\documentclass[reprint,
 amsmath,amssymb,
 aps,
prstab
]{revtex4-1}

\usepackage{graphicx}% Include Figure files
\usepackage{subcaption}
\usepackage{dcolumn}% Align table columns on decimal point
\usepackage{bm}% bold math
\usepackage{xcolor}
\usepackage{natbib} 
\usepackage[nottoc,numbib]{tocbibind}
\usepackage{amsmath}
\usepackage{lipsum}

\usepackage[%
  colorlinks=true,
  urlcolor=blue,
  linkcolor=blue,
  citecolor=blue
]{hyperref}

\usepackage{stackengine}
\usepackage{overpic}% super impose on the picture
\usepackage{pict2e}
\usepackage{tikz}

\usepackage{amsmath}

\usepackage{tabularx}
\usepackage{xcolor}

\usepackage{textcomp} %Textmu

\makeatletter
\newcommand\thefontsize[1]{{#1 The current font size is: \f@size pt\par}}
\newcommand\thefontsizeHere{{The current font size is: \f@size pt\par}}
\makeatother

\usepackage{tikz,xcolor,hyperref}

\definecolor{lime}{HTML}{A6CE39}
\DeclareRobustCommand{\orcidicon}{%
	\begin{tikzpicture}
	\draw[lime, fill=lime] (0,0) 
	circle [radius=0.16] 
	node[white] {{\fontfamily{qag}\selectfont \tiny ID}};
	\draw[white, fill=white] (-0.0625,0.095) 
	circle [radius=0.007];
	\end{tikzpicture}
	\hspace{-2mm}
}

\foreach \x in {A, ..., Z}{%
	\expandafter\xdef\csname orcid\x\endcsname{\noexpand\href{https://orcid.org/\csname orcidauthor\x\endcsname}{\noexpand\orcidicon}}
}

\begin{document}

\preprint{APS/123-QED}

\title{Beam Breakup Instability Studies of \\Powerful Energy Recovery Linac for Experiments}% Force line breaks with \\
%\title{An investigation of RF instabilities during beam loading of recirculating Energy Recovery Linac}% Force line breaks with \\
% Author Orchid ID: enter ID or remove command
\newcommand{\orcidauthorA}{0000-0002-5903-8930} % Sadiq; Add \orcidA{} behind the author's name
\newcommand{\orcidauthorB}{0000-0001-6346-5989} % Rob; Add \orcidA{} behind the author's name
\newcommand{\orcidauthorC}{0000-0002-8987-4999} % Peter; Add \orcidA{} behind the author's name
\newcommand{\orcidauthorD}{0000-0001-7594-5840} % Carmelo; Add \orcidA{} behind the author's name
\newcommand{\orcidauthorE}{0000-0001-8705-9539} % Ryan; Add \orcidE{} behind the author's name
\newcommand{\orcidauthorF}{0000-0003-2028-4330} % Kirsten; Add \orcidE{} behind the author's name
\newcommand{\orcidauthorG}{0000-0003-1249-5293} % Kirsten; Add \orcidG{} behind the author's name

\author{Sadiq~Setiniyaz\orcidA{}}
\email{saitiniy@jlab.org} 
\thanks{Now at Center for Advanced Studies of Accelerators, Jefferson Lab, Newport News, USA}
%\affiliation{Current: Center for Advanced Studies of Accelerators, Jefferson Lab, Newport News, USA}%
%\affiliation{Previous: Engineering Department, Lancaster University, Lancaster, LA1 4YW, UK }
%\affiliation{Previous: Cockcroft Institute, Daresbury Laboratory, Warrington, WA4 4AD, UK}

\author{R.~Apsimon\orcidB{}}
\email{r.apsimon@lancaster.ac.uk}
\affiliation{Engineering Department, Lancaster University, Lancaster, LA1 4YW, UK }
\affiliation{Cockcroft Institute, Daresbury Laboratory, Warrington, WA4 4AD, UK}

\author{P.~H.~Williams\orcidC{}}
%\email{peter.williams@stfc.ac.uk}
\affiliation{STFC Daresbury Laboratory \& Cockcroft Institute, Warrington, WA4 4AD, UK}

\author{C.~Barbagallo\orcidD{}}
\affiliation{Laboratoire de Physique des 2 Infinis Irène Joliot-Curie (IJCLab), Orsay, France}
\affiliation{The Paris-Saclay University, Gif-sur-Yvette, France}
\thanks{Now at CERN, Geneva, Switzerland}

\author{S.~A.~Bogacz\orcidG{}}
\affiliation{Center for Advanced Studies of Accelerators, Jefferson Lab, Newport News, USA}%

\author{R.~M.~Bodenstein\orcidE{}}
\affiliation{Center for Advanced Studies of Accelerators, Jefferson Lab, Newport News, USA}%

\author{K.~Deitrick\orcidF{}}
\affiliation{Center for Advanced Studies of Accelerators, Jefferson Lab, Newport News, USA}%

\date{\today}% It is always \today, today,

\begin{abstract}
The maximum achievable beam current in an Energy Recovery Linac (ERL) is often constrained by Beam Breakup (BBU) instability. Our previous research highlighted that filling patterns have a substantial impact on BBU instabilities in multi-pass ERLs.
In this study, we extend our investigation to the 8-cavity model of the Powerful ERL for Experiment (PERLE). We evaluate its requirements for damping cavity Higher Order Modes (HOMs) and propose optimal filling patterns and bunch timing strategies.
Our findings reveal a significant new insight: while filling patterns are crucial, the timing of bunches also plays a critical role in mitigating HOM beam loading and BBU instability. This previously underestimated factor is essential for effective BBU control. 
We estimated the PERLE threshold current using both analytical and numerical models, incorporating the designed PERLE HOM dampers. During manufacturing, HOM frequencies are expected to vary slightly, with an assumed RMS frequency jitter of 0.001 between cavities for the same HOM. Introducing this jitter into our models, we found that the dampers effectively suppressed BBU instability, achieving a threshold current an order of magnitude higher than the design requirement.
Our results offer new insights into ERL BBU beam dynamics and have important implications for the design of future ERLs.
\end{abstract}

\maketitle

%\begin{itemize}
%\item short intro into recirculating ERLs and PERLE
%\subitem citing design studies around the world
%\subitem PERLE description
%\item BBU theoretical model
%\subitem multiple checkpoint model 
%\item Allowed patterns for PERLE
%\item BBU simulation
%\subitem explain code
%\subitem results
%\item Conclusion
%\item acknowledgments
%\end{itemize}

\section{Introduction}

In the 2020 European Strategy for Particle Physics~\cite{EuropeanStrategyGroup:2020pow}, research and development in the field of superconducting Energy Recovery Linacs (ERLs)~\cite{Merminga2016} was prioritized, owing to their anticipated crucial role in future particle physics applications. The PERLE (Powerful ERL Experiment)~\cite{Angal2018perle} is one of the suggested advanced ERL test facilities developed to assess potential options for a 50 GeV ERL, as proposed in the Large Hadron Electron Collider (LHeC)~\cite{Brüning2022, Klein2018, Agostini2021} and Future Circular Collider (FCC-eh) designs. Moreover, it functions as a base for dedicated experiments in nuclear and particle physics. PERLE's main objective is to investigate the operation of high current, continuous wave (CW), multi-pass systems utilizing superconducting cavities that operate at 802 MHz.

PERLE's remarkable capacity to handle beam power up to 10~MW and an operating (injection) current of 20~mA offers researchers  a unique opportunity to conduct controlled studies on Beam BreakUp (BBU)~\cite{LYNEIS1983269} studies relevant to next-generation multi-pass ERL designs. The maximum achievable beam current in an ERL is often constrained by the BBU instability. Recent research has demonstrated that the selection of filling patterns, which delineates the order of bunch injection into the ERL over subsequent turns, can significantly influence not only RF stability and cavity voltage~\cite{Setiniyaz2020} but also BBU instabilities in multi-pass ERLs~\cite{Setiniyaz2021}. However, the impact on BBU is somewhat more complex due to the asynchronous nature of the HOM mode relative to the beam, and the transcendental relationship between the HOM voltage and the bunch offsets. 

Filling patterns describe the order bunches pass through the cavity in multi-turn ERLs. For example, the simplest filling pattern [1~2~3~4~5~6] describes a pattern where first bunch is at its first turn, the second bunch is at its second turn so on so forth. Another filling pattern [1~4~2~5~3~6] describes a pattern where first bunch is at its first turn, the second bunch is at its fourth turn, and so on. When the filling pattern does not change when bunches pass though the cavity, this pattern is referred as ``Sequence Preserving'' (SP) patterns. More complicated filling patterns can be achieved via maneuvering bunch injection timing and re-combination schemes, which is beyond scope of this paper. In this paper, we shall only focus on the SP patterns.

\begin{figure}
\scalebox{0.35} [0.35]{\includegraphics[trim={2.0cm 3.6cm 2.0cm 4.1cm},clip]{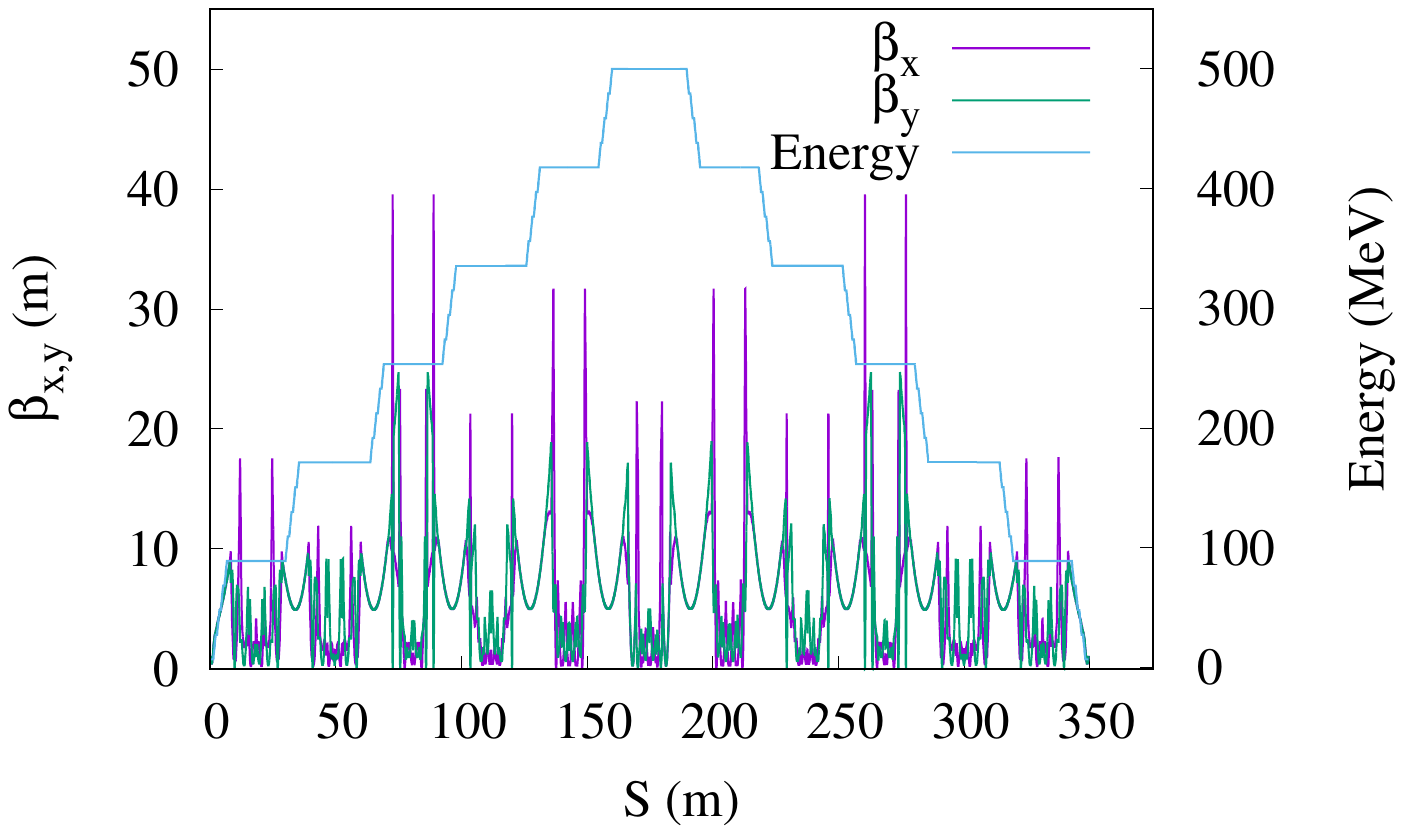}}
	\caption{Complete multi-pass PERLE optics (3-passes `up' and 3 passes `down')}
	\label{fig:Optics} 
\end{figure}

The original optics for PERLE 2.0~\cite{PERLE} and subsequent PERLE 2.1~\cite{michaud} features symmetric, by-design, multi-pass linacs, with minimized values of beta functions reaching about 10 meters at both linac ends. An important design choice was to maintain almost identical optics for all accelerating and decelerating passes (except for the first pass), as shown in Fig.~\ref{fig:Optics}.

%\begin{figure}
%\scalebox{0.7} [0.7]{\includegraphics{fig/multi-pass_500MeV.png}}
%	\caption{Complete multi-pass PERLE 2.0 optics (3-passes 'up' and 3 passes 'down')}
%	\label{fig:Optics} 
%\end{figure}

Since PERLE is a 6-pass ERL, the current pass through the cavity during the operation is 6 times that of the injected current. Therefore, for 20~mA injection, the current passes through the cavities at 0.12~Amps. Note the threshold current in this paper refers to the current pass through the cavity.

\section{Multiple checkpoint analytical model}
The numerical simulation will be benchmarked against the multiple checkpoint analytical model, which was previously discussed in~\cite{yunn2005, Carmelo2022ERL, Bogacz_PERLE}. The threshold current $I_{\mathrm{th, \lambda}}$ of the mode number $\lambda$ with angular frequency $\omega_{\lambda}$ is given by:

\begin{equation} 
    I_{\mathrm{th, \lambda}} = \frac{-2E}{e\left(\frac{R}{Q}\right)_{\lambda}Q_{L, \lambda}k_{\lambda}\sum_{j>i=1}^{N_{c}}\left(\frac{E}{E_{j}}\right)(M^{ij})_{mn}\sin{(\omega_{\lambda} t_{r}^{ij})}}.
    \label{eq:IthAna}
\end{equation}

In this equation:
\begin{itemize}
    \item $e$  is the electron charge.
    \item $k_{\lambda}$ is the wave number of the HOM.
    \item $E$  is the energy of the beam in the recirculation arc.
    \item \( E_{j} \) denotes the beam energy at checkpoint \( j \).
    \item \( (M^{ij})_{mn} \) is an element of \( M^{ij} \), the transfer matrix from the \( i^{th} \) checkpoint to the \( j^{th} \) checkpoint; mn equals to 12 for horizontal modes and 34 for vertical modes.
    \item \( t_{r}^{ij} \) is the time for particle travel between corresponding checkpoints.
    \item \( \left(\frac{R}{Q}\right)_{\lambda} \) is the shunt impedance of the HOM.
    \item \( Q_{L, \lambda} \) is the loaded quality factor of the HOM.
    \item \( N_{c} \) represents the total number of checkpoints, specifically positioned at the exits of the linacs.
\end{itemize}

%To streamline the computational analysis, we have set \(\sin{(\omega_{\lambda} t_{r}^{ij})} = 1\), representing the maximal attainable value for this trigonometric function. This assumption yields the minimal possible threshold current, thereby providing a conservative estimate in our theoretical framework. 
It is crucial to note that this analytical model makes certain simplifications, particularly overlooking the impact of the filling pattern and bunch timing on the calculated threshold current. As highlighted in Ref.~\cite{Setiniyaz2021}, the bunch filling pattern exerts a significant influence on both the BBU and the threshold current. 
Moreover, bunch timing is instrumental in determining the HOM arrival phase for succeeding particles and also has big impact on BBU, which is elaborated upon in subsequent sections. Therefore, while the numerical and analytical models may not precisely coincide, a general level of agreement between them can still be expected.

\section{Constraints for PERLE filling patterns}
\label{sec:PattConstraints}
In a multi-turn ERL, as discussed in previous studies \cite{Setiniyaz2020, Setiniyaz2021}, depending on the topology of the beam line, there can be 1, $N$/2 or $N$ arcs on each side of the ring; where $N$ is the number of turns each bunch completes in the ERL between injection and extraction. In many situations, the preferred option is to opt for $N$/2 arcs as this avoids the inherent complexity of an FFA-type arc design with a very high energy acceptance, while also reducing the complexity and capital cost associated with N arcs on each side of the ring. In the case of PERLE, there are several key constraints placed on the injection filling pattern of the bunches in the ring:
\begin{itemize}
    \item Bunches should be as close to evenly spaced as possible to minimize collective effects
    \item The beam line should be configured for a sequence preserving (SP) scheme, in order to ensure regular injection timing
\end{itemize}

From the first constraint, this implies that we need a filling pattern such that accelerating and decelerating bunches alternate. From previous work \cite{Setiniyaz2020, Setiniyaz2021}, it is known that every filling pattern has an associated SP transition set. A filling pattern is defined as the RF cycle or bucket that is occupied by the bunch on each turn as is represented by a row vector, where the index is the turn number and the value is the RF cycle/bucket number. A transition set is a row vector that shows how many RF cycles/buckets a bunch on turn $j$ shifts when it starts turn $\left(j+1\right)$. For a sequence preserving scheme, we require that on each turn, the bunch on it’s $j^{\text{th}}$ turn occupies RF cycle number $F_{j}$. Therefore, after each turn, the bunch on turn j shifts from RF cycle $F_{j}$ to $F_{\left(j+1\right)}$, therefore the filling pattern and transition set for an SP scheme are related as:

\begin{equation}
    \begin{array}{l}
    F=\left[F_{1}~F_{2}~\cdots~F_{N}\right] \\
    T= \{\left(F_{2}-F_{1}\right)~\left(F_{3}-F_{2}\right)~\cdots~\left(F_{N}-F_{1}\right)\}
    \end{array}
\end{equation}

PERLE is a 6-turn ERL and the bunch train is 20 RF cycles long. Usually, we would define the filling pattern as the RF bucket occupied by a bunch on turn $j$, however in this case, it is more useful to define it as the RF cycle modulo 20. It is also useful to note that without loss of generality, we can define that the bunch on turn 1 is in RF cycle 1. For the PERLE filling pattern, there are a total of 12 filling patterns out of 120 unique patterns which meet the constraints. These are summarized in Table~\ref{tab:allowedFP1}.
%$\left[1~3~5~2~4~6\right]$~$\left[1~3~5~2~6~4\right]$~$\left[1~3~5~4~2~6\right]$~$\left[1~3~5~4~6~2\right]$ \\
%$\left[1~3~5~6~2~4\right]$~$\left[1~3~5~6~4~2\right]$~$\left[1~5~3~2~4~6\right]$~$\left[1~5~3~2~6~4\right]$ \\
%$\left[1~5~3~4~2~6\right]$~$\left[1~5~3~4~6~2\right]$~$\left[1~5~3~6~2~4\right]$~$\left[1~5~3~6~4~2\right]$ \\

\begin{table}
	\caption{Allowed filling patterns of PERLE.}
	\begin{ruledtabular}
		\begin{tabular}{cccc}
$\left[1~3~5~2~4~6\right]$&$\left[1~3~5~2~6~4\right]$&$\left[1~3~5~4~2~6\right]$&$\left[1~3~5~4~6~2\right]$ \\
$\left[1~3~5~6~2~4\right]$&$\left[1~3~5~6~4~2\right]$&$\left[1~5~3~2~4~6\right]$&$\left[1~5~3~2~6~4\right]$ \\
$\left[1~5~3~4~2~6\right]$&$\left[1~5~3~4~6~2\right]$&$\left[1~5~3~6~2~4\right]$&$\left[1~5~3~6~4~2\right]$ \\
		\end{tabular}
	\end{ruledtabular}
	\label{tab:allowedFP1}
\end{table}

\begin{figure}
\scalebox{0.22} [0.22]{\includegraphics{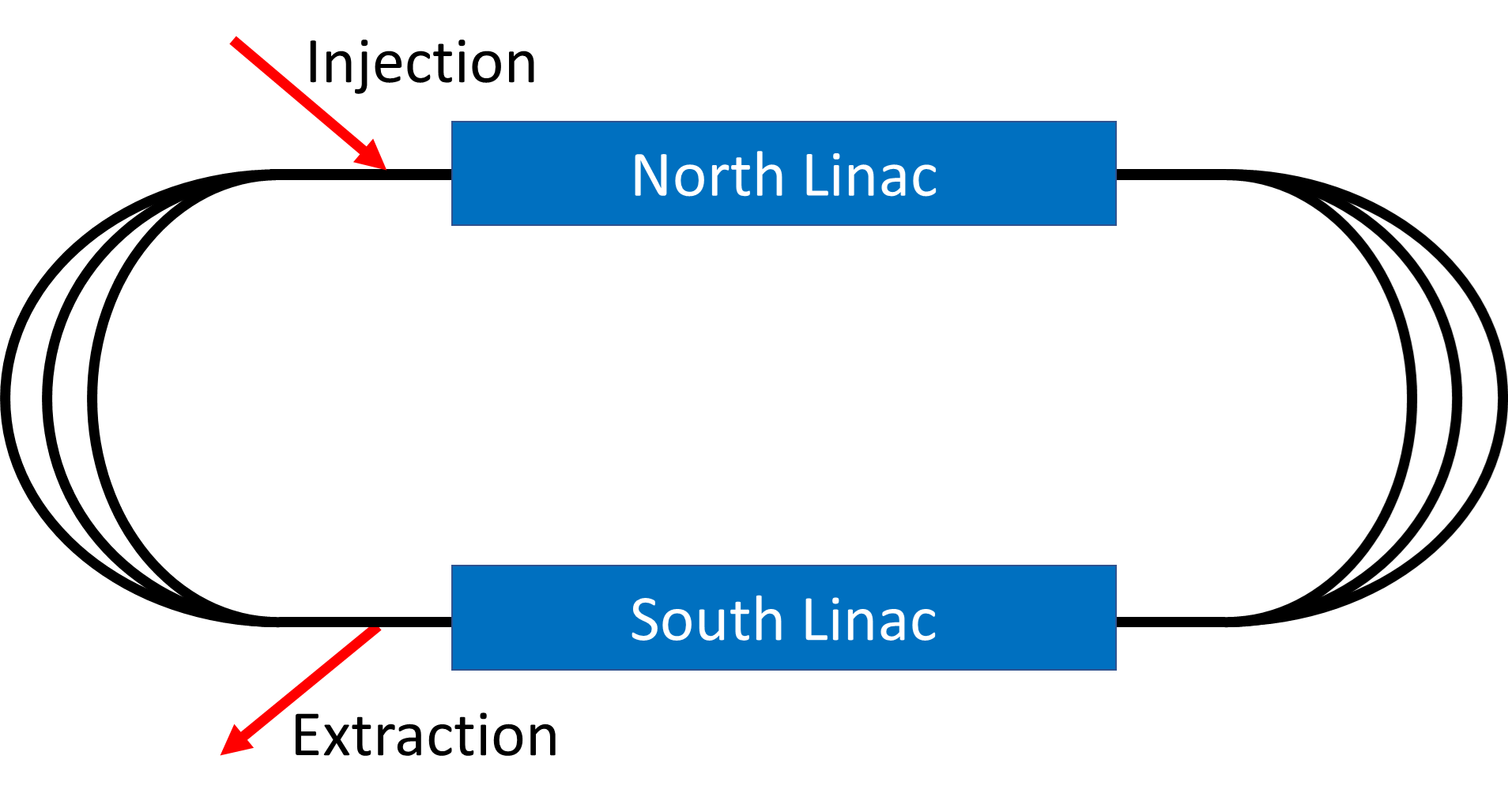}}
	\caption{PERLE layout diagram. }
	\label{fig:diagram} 
\end{figure}

Given the bijective relationship that this has to the SP transition sets, we essentially know the total length of each turn modulo 20. From this, we will infer the arc lengths. To begin with, we shall define a few conventions. The beam is injected in the North straight section, just upstream of the North linac. The beam travels in a clockwise direction, passing through the East arc section, followed by the South straight section with the South linac, and finally the West arc section to complete a full turn of the ring. The beam is extracted in the South straight section, just downstream of the South linac  as shown in the diagram in Fig.~\ref{fig:diagram}. The straight sections are assumed to be an equal length, $L$, and the arc lengths are defined as $A_{n}$, where $A_{1}$, $A_{3}$ and $A_{5}$ are the East arcs and $A_{2}$, $A_{4}$ and $A_{6}$ are the West arcs. $A_{0}$ is a time delay between extraction of the bunch on turn 6 and the injection of a new bunch. The lengths of each turn, with respect to the injection point in the North straight is given as:

\begin{equation}
    \begin{array}{l}
    T_{N1} = 2L + A_{1} + A_{2} = 20m_{N1} + F_{N2} - F_{N1} \\
    T_{N2} = 2L + A_{3} + A_{4} = 20m_{N2} + F_{N3} - F_{N2} \\
    T_{N3} = 2L + A_{5} + A_{6} = 20m_{N3} + F_{N4} - F_{N3} \\
    T_{N4} = 2L + A_{5} + A_{4} = 20m_{N4} + F_{N5} - F_{N4} \\
    T_{N5} = 2L + A_{3} + A_{2} = 20m_{N5} + F_{N6} - F_{N5} \\
    T_{N6} = 2L + A_{1} + A_{0} = 20m_{N6} + F_{N1} - F_{N6}
    \end{array}
\end{equation}

The denotation of $F_{Ni}$ is to represent the filling pattern in the North straight. If we assume that we know the length of the straights and we define one arc length, then we can define all other arc lengths in terms of it. In this example, we will assume that the length of Arc 6 is known:

\begin{equation}
    \begin{array}{l}
    A_{0} \equiv\left(A_{6}+2F_{N1}-2F_{N4}\right)\text{mod}\left(20\right) \\
    A_{1} \equiv\left(-2L-A_{6}-F_{N1}-F_{N6}+2F_{N4}\right)\text{mod}\left(20\right) \\
    A_{2} \equiv\left(A_{6}+F_{N2}+F_{N6}-2F_{N4}\right)\text{mod}\left(20\right) \\
    A_{3} \equiv\left(-2L-A_{6}-F_{N2}-F_{N5}+2F_{N4}\right)\text{mod}\left(20\right) \\
    A_{4} \equiv\left(A_{6}+F_{N3}+F_{N5}-2F_{N4}\right)\text{mod}\left(20\right) \\
    A_{5} \equiv\left(-2L-A_{6}+F_{N4}-F_{N3}\right)\text{mod}\left(20\right)
    \end{array}
\end{equation}

Having determined the arc lengths, we can now look at the resulting filling patterns in the South straight as this is the North filling pattern, plus a straight length and an arc length:

\begin{equation}
    \begin{array}{l}
    F_{S1} = F_{N1} + L + A_{1} \\
    F_{S2} = F_{N2} + L + A_{3} \\
    F_{S3} = F_{N3} + L + A_{5} \\
    F_{S4} = F_{N4} + L + A_{5} \\
    F_{S5} = F_{N5} + L + A_{3} \\
    F_{S6} = F_{N6} + L + A_{1}
    \end{array}
\end{equation}

We can now substitute in the relevant arc lengths to obtain:
\begin{equation}
    \begin{array}{l}
    F_{S1} = \left(2F_{N4} - L - A_{6}\right) - F_{N6} \\
    F_{S2} = \left(2F_{N4} - L - A_{6}\right) - F_{N5} \\
    F_{S3} = \left(2F_{N4} - L - A_{6}\right) - F_{N4} \\
    F_{S4} = \left(2F_{N4} - L - A_{6}\right) - F_{N3} \\
    F_{S5} = \left(2F_{N4} - L - A_{6}\right) - F_{N2} \\
    F_{S6} = \left(2F_{N4} - L - A_{6}\right) - F_{N1}
    \end{array}
\end{equation}

However, we can add or subtract a constant from the filling pattern as this is simply equivalent to reindexing the RF cycle numbers, and therefore we obtain that the North and South filling patterns must be related as $F_{S}\equiv-F_{N}^{m}$, where $F^{m}$ is used to denote that the order of the filling pattern elements are flipped. This result shows that in general the filling pattern in the north and south arcs are generally different.

%\textcolor{red}{Previous studies have been undertaken and clearly demonstrate the importance of the injection timing sequence (known as the filling pattern) as well as the recirculation scheme or beam line topology \cite{Setiniyaz2020, Setiniyaz2021}. The PERLE baseline design comprises of a 6-turn ERL, with 3 arcs at either end of the ERL ring, with bunches injected every 20 RF cycles.}

%    PattIdxN0 = [51, 53,  57, 59, 75, 77,  81, 83, 99, 101, 105, 107]; % 12 patterns
%    PattIdxS0 = [51, 105, 81, 59, 75, 101, 57, 83, 99, 77,  53,  107]; % 12 patterns
%\begin{table}
%	\caption{Patterns in the north and south linacs.}
%	\begin{ruledtabular}
%		\begin{tabular}{ccccccccccccc}
%		 north  & 51& 53&  57& 59& 75& 77&  81& 83& 99& 101& 105& 107\\
%		 south  & 51& 105& 81& 59& 75& 101& 57& 83& 99& 77&  53&  107\\
%		\end{tabular}
%	\end{ruledtabular}
%	\label{tab:Iths}
%\end{table}

\section{Bunch timing and Filling patterns for PERLE}

\begin{figure}
\scalebox{0.4} [0.4]{\includegraphics{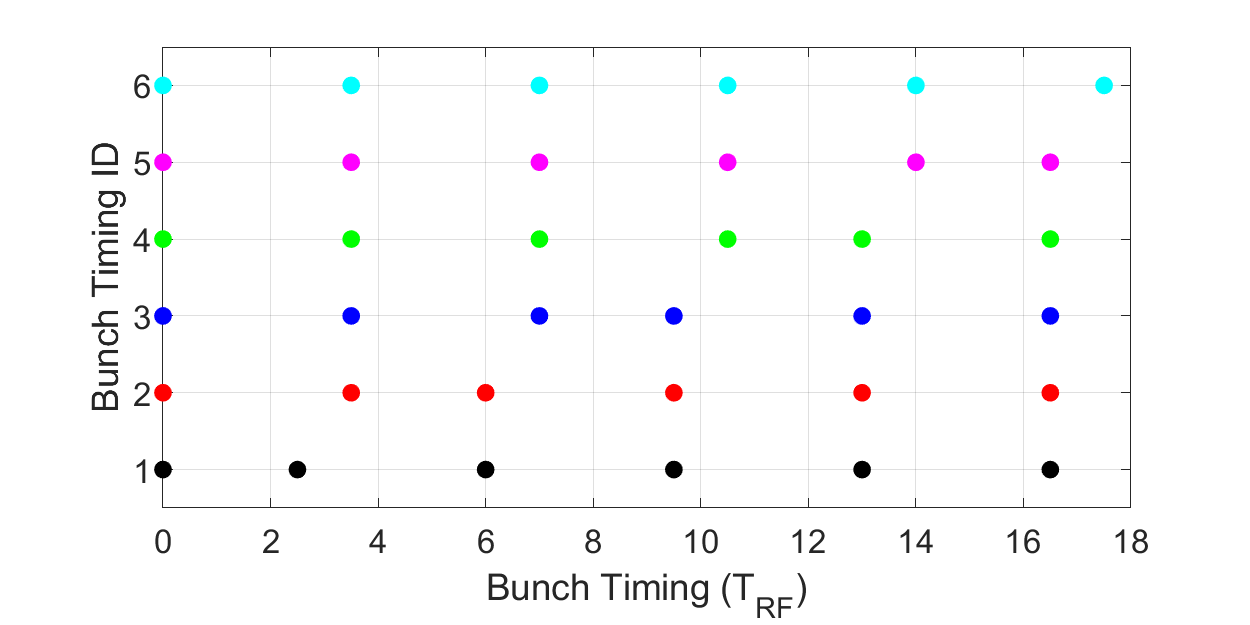}}
	\caption{Summary of 6 bunch timing combinations (tc). }
	\label{fig:timings} 
\end{figure}

We have examined six different bunch timings, which are detailed in Fig.~\ref{fig:timings}. Each timing ID represents a unique pattern of bunch spacing in terms of the fundamental RF period \( T_{\text{RF}} \), approximately \(1.25 \, \text{ns}\). For instance, in Timing ID 1, the first bunch occurs at \( T_{\text{RF}} = 0 \), and the next bunch appears \(2.5 \, T_{\text{RF}}\) later. Integer multiples of \( T_{\text{RF}} \) indicate acceleration, while half-integer multiples suggest deceleration. A single bunch train spans \(0 - 20 \, T_{\text{RF}}\), and the subsequent train begins at \(20 \, T_{\text{RF}}\). If we put two trains together, for example in case of timing ID 1, it would have timing of [0~~2.5~~6~~9.5~~13~~16.5~~20~~22.5~~26~~29.5~~33~~36.5].

%\begin{table}
%	\caption{Summary of Bunch Timing Patterns. \textcolor{red}{(This is the wrong convention based on the maths, but the other convention can't be written out this simply, we need to think about how to explain this as it's not trivial!)}}
%	\begin{ruledtabular}
%		\begin{tabular}{lc}
%			Timing ID & Timing \([T_{\text{RF}}]\) \\
%			\hline
%			1 & \([0, 2.5, 6, 9.5, 13, 16.5]\) \\
%			2 & \([0, 3.5, 6, 9.5, 13, 16.5]\) \\
%			3 & \([0, 3.5, 7, 9.5, 13, 16.5]\) \\
%			4 & \([0, 3.5, 7, 10.5, 13, 16.5]\) \\
%			5 & \([0, 3.5, 7, 10.5, 14, 16.5]\) \\
%			6 & \([0, 3.5, 7, 10.5, 14, 17.5]\) \\
%		\end{tabular}
%	\end{ruledtabular}
%	\label{tab:timings}
%\end{table}

Table~\ref{tab:ptc} enumerates the feasible combinations of timing and filling patterns for the North and South linacs. In each entry, the first number represents the timing ID (or filling pattern) for the North linac, and the second pertains to the South linac. It is notable that the timing often differs between the North and South linacs, resulting in variable relative bunch spacing.

\begin{table*}
	\caption{Permissible Combinations of Timing and Filling Patterns in North and South Linacs.}
	\begin{ruledtabular}
		\begin{tabular}{lcccccccccccc}
			& [N, S] & [N, S] & [N, S] & [N, S] & [N, S] & [N, S] & [N, S] & [N, S] & [N, S] & [N, S] & [N, S] & [N, S] \\
			\hline
			Pattern & [51, 51] & [53, 105] & [57, 81] & [59, 59] & [75, 75] & [77, 101] & [81, 57] & [83, 83] & [99, 99] & [101, 77] & [105, 53] & [107, 107] \\
			\hline
			Timing 1 & [1, 5] & [1, 3] & [1, 5] & [1, 1] & [1, 3] & [1, 1] & [1, 5] & [1, 3] & [1, 5] & [1, 1] & [1, 3] & [1, 1] \\
			Timing 2 & [2, 4] & [2, 2] & [2, 4] & [2, 6] & [2, 2] & [2, 6] & [2, 4] & [2, 2] & [2, 4] & [2, 6] & [2, 2] & [2, 6] \\
			Timing 3 & [3, 3] & [3, 1] & [3, 3] & [3, 5] & [3, 1] & [3, 5] & [3, 3] & [3, 1] & [3, 3] & [3, 5] & [3, 1] & [3, 5] \\
			Timing 4 & [4, 2] & [4, 6] & [4, 2] & [4, 4] & [4, 6] & [4, 4] & [4, 2] & [4, 6] & [4, 2] & [4, 4] & [4, 6] & [4, 4] \\
			Timing 5 & [5, 1] & [5, 5] & [5, 1] & [5, 3] & [5, 5] & [5, 3] & [5, 1] & [5, 5] & [5, 1] & [5, 3] & [5, 5] & [5, 3] \\
			Timing 6 & [6, 6] & [6, 4] & [6, 6] & [6, 2] & [6, 4] & [6, 2] & [6, 6] & [6, 4] & [6, 6] & [6, 2] & [6, 4] & [6, 2] \\
		\end{tabular}
	\end{ruledtabular}
	\label{tab:ptc}
\end{table*}

Out of 12 permissible filling pattern combinations, six exhibit identical patterns in both North and South linacs, while the remaining six differ. Each pattern combination comprises two numbers: the filling pattern in the North linac and that in the South linac. For conciseness, we refer to 120 filling patterns in the 6-turn ERL by their filling pattern number, which is given as follows:

\begin{equation}
\begin{split}
    \text{Pattern 1: } &[1, 2, 3, 4, 5, 6], \\
    \text{Pattern 2: } &[1, 2, 3, 4, 6, 5], \\
    &\vdots \\
    \text{Pattern 120: } &[1, 6, 5, 4, 3, 2].
\end{split}
\label{eq:ustored0}
\end{equation}

It is crucial to distinguish between two conventions for describing filling patterns: the \textit{space convention} and the \textit{time convention}.

\begin{itemize}
    \item \textbf{Space Convention}: In this convention, the numerical value represents the turn number, and its position within the array (i.e., its index) indicates its physical location in the bunch train (i.e., its RF bucket number).
    \item \textbf{Time Convention}: Conversely, in this convention, the numerical value signifies its physical order within the bunch train (i.e., its RF bucket number), and its position (i.e., its index) represents the turn number.
\end{itemize}

To put it simply, if the index corresponds to the physical location in the bunch train, then the ``space convention'' is being used. If the index denotes the turn number, then the ``time convention'' applies.

For instance, consider a filling pattern described in the time convention:

\begin{equation}
[1~3~5~2~4~6] = [1_{\textcolor{blue}{1}}~3_{\textcolor{blue}{2}}~5_{\textcolor{blue}{3}}~2_{\textcolor{blue}{4}}~4_{\textcolor{blue}{5}}~6_{\textcolor{blue}{6}}].
\label{eq:fpcon_time}
\end{equation}

This filling pattern can be translated into the space convention as:

\begin{equation}
[\textcolor{blue}{1}_{1}~\textcolor{blue}{4}_{2}~\textcolor{blue}{2}_{3}~\textcolor{blue}{5}_{4}~\textcolor{blue}{3}_{5}~\textcolor{blue}{6}_{6}] = [\textcolor{blue}{1~4~2~5~3~6}]
\label{eq:fpcon_space}
\end{equation}

In this example, the index shifts between the two conventions, reflecting either its physical location (space convention) or its turn number (time convention).

The space convention is often favored for its intuitive grasp when describing filling patterns, whereas the time convention is particularly useful for computational tasks such as pattern transitions and arc length calculations. In this paper, we restrict the usage of the time convention to the section outlined in \ref{sec:PattConstraints}. It should be noted that the filling patterns provided in Table \ref{tab:allowedFP1} are expressed in the time convention. These have been subsequently translated to the more intuitive space convention, as presented in Table \ref{tab:allowedFP2}.

\begin{table}
	\caption{Conversion between filling pattern conventions.}
	\begin{ruledtabular}
		\begin{tabular}{ccc}
	time  & space  & pattern number in\\
 convention &  convention &  space convention\\
 \hline
$\left[1~3~5~2~4~6\right]$&[1~4~2~5~3~6] & 51 \\
$\left[1~3~5~2~6~4\right]$&[1 4 2 6 3 5] & 53 \\
$\left[1~3~5~4~2~6\right]$&[1 5 2 4 3 6] & 75\\
$\left[1~3~5~4~6~2\right]$&[1 6 2 4 3 5] &  99\\
$\left[1~3~5~6~2~4\right]$&[1 5 2 6 3 4] & 77\\
$\left[1~3~5~6~4~2\right]$&[1 6 2 5 3 4] & 101 \\
$\left[1~5~3~2~4~6\right]$&[1 4 3 5 2 6] & 57 \\
$\left[1~5~3~2~6~4\right]$&[1 4 3 6 2 5] & 60 \\
$\left[1~5~3~4~2~6\right]$&[1 5 3 4 2 6] & 81 \\
$\left[1~5~3~4~6~2\right]$&[1 6 3 4 2 5] & 105 \\
$\left[1~5~3~6~2~4\right]$&[1 5 3 6 2 4] & 83 \\
$\left[1~5~3~6~4~2\right]$&[1 6 3 5 2 4] & 107 \\
		\end{tabular}
	\end{ruledtabular}
	\label{tab:allowedFP2}
\end{table}

\section{Higher Order Dipole Modes}

The PERLE cavity is a 5-cell superconducting cavity that operates at a fundamental mode frequency of 801.58 MHz. The dipole Higher Order Modes (HOMs) deemed most critical for the PERLE bare cavity are itemized in Table~\ref{tab:PERLEHOM}~\cite{Carmelo2022ERL}. Dipole modes are known to introduce transverse kicks to the beam, whereas monopole modes are responsible for inducing energy jitter. However, the influence of monopole modes is largely inconsequential, owing to their markedly diminished amplitude relative to the fundamental mode. Furthermore, any resultant energy jitter can be effectively mitigated through other compensatory methods. In light of the more significant consequences of dipole modes, this paper will specifically concentrate on evaluating their criticality and impact.

\begin{table}
	\caption{PERLE HOMs.}
	\begin{ruledtabular}
		\begin{tabular}{rccc}
	HOM & mode & frequency & {R/Q}\textsubscript{t}\\
            &      &    [GHz]        &      [\textohm]\\
			\hline
1& TE\textsubscript{111} $\pi/5$ & 0.9342 & 0.2341 \\
2& TE\textsubscript{111} $2\pi/5$ & 0.9597 & 0.0037\\
3& TE\textsubscript{111} $3\pi/5$ & 0.9964 & 28.3557\\
4& TE\textsubscript{111} $4\pi/5$ & 1.0365 & 61.0159\\
5& TE\textsubscript{111} $\pi$ & 1.0768 & 12.3017 \\
6& TM\textsubscript{110} $\pi$ & 1.0965 & 0.9020\\
7& TM\textsubscript{110} $4\pi/5$ & 1.1255 & 30.9362 \\
8& TM\textsubscript{110} $3\pi/5$ & 1.1471 & 37.1876\\
9& TM\textsubscript{110} $2\pi/5$ & 1.1592 & 6.3241\\
10& TM\textsubscript{110} $\pi/5$ & 1.1629 & 0.4378\\
11& TM\textsubscript{111} $\pi$ & 1.4662 & 0.1575\\
12& TM\textsubscript{111} $4\pi/5$ & 1.4839 & 5.5573\\
13& TM\textsubscript{111} $3\pi/5$ & 1.4997 & 4.0800\\
14& TM\textsubscript{111} $2\pi/5$ & 1.5012 & 16.8400\\
15& TM\textsubscript{111} $\pi/5$ & 1.5532 & 9.8024\\ 
		\end{tabular}
	\end{ruledtabular}
	\label{tab:PERLEHOM}
\end{table}

%\subsection{\textcolor{red}{8-cavity mode}l}

\section{Critical HOMs $Q_L$ Simulations}

The impact filling pattern on the BBU hasn't been investigated previously and a code  with filling pattern capability hasn't developed yet. Hence, we have developed a BBU code described in the Ref.~\cite{Bogacz_PERLE}. The current code is an extension of the single-cavity and single-mode model described Ref.~\cite{Setiniyaz2021} to a multi-cavity and multi-mode model. During the development process, the ERLBBU algorithm in Refs.~\cite{Pozdeyev2005, TennantDissertation2006, Pozdeyev2006} was adapted, and extended by adding filling pattern and timing dependence. 

\subsection{Simulation and analytical results}
The first step of the BBU instability studies are to estimate the required critical quality factor $Q_L$ of HOMs to operate the PERLE at 0.12 Amps. The simulation results for the 15 HOMs are given in Fig.~\ref{fig:SimRes1}. The black dashed lines are the estimation by the analytical model, while the colored lines are from simulation results by using different timing combinations given in Table~\ref{tab:ptc}. The maximum $Q_L$ value for simulation is set to 10$^{10}$, as this is near the $Q_L$ value of the fundamental mode, making further simulation redundant. 

Firstly, we see the analytical model and simulation exhibit good agreement, as evidenced by the close proximity of the dashed black lines to the colored lines. However, unlike the analytical model, the simulation can capture the filling pattern and bunch timing dependence of the BBU instability, resulting in a more accurate prediction. Secondly, some modes are more sensitive to pattern and filling combinations than others. Higher frequency HOMs are less sensitive to the timing combinations than the lower frequency HOMs. Lastly, both pattern and timing combination can vary the $Q_L$ by an order of magnitude or more, which indicates they both are critical for BBU instability suppression.   

\begin{figure*}
\scalebox{0.42} [0.42]{\includegraphics{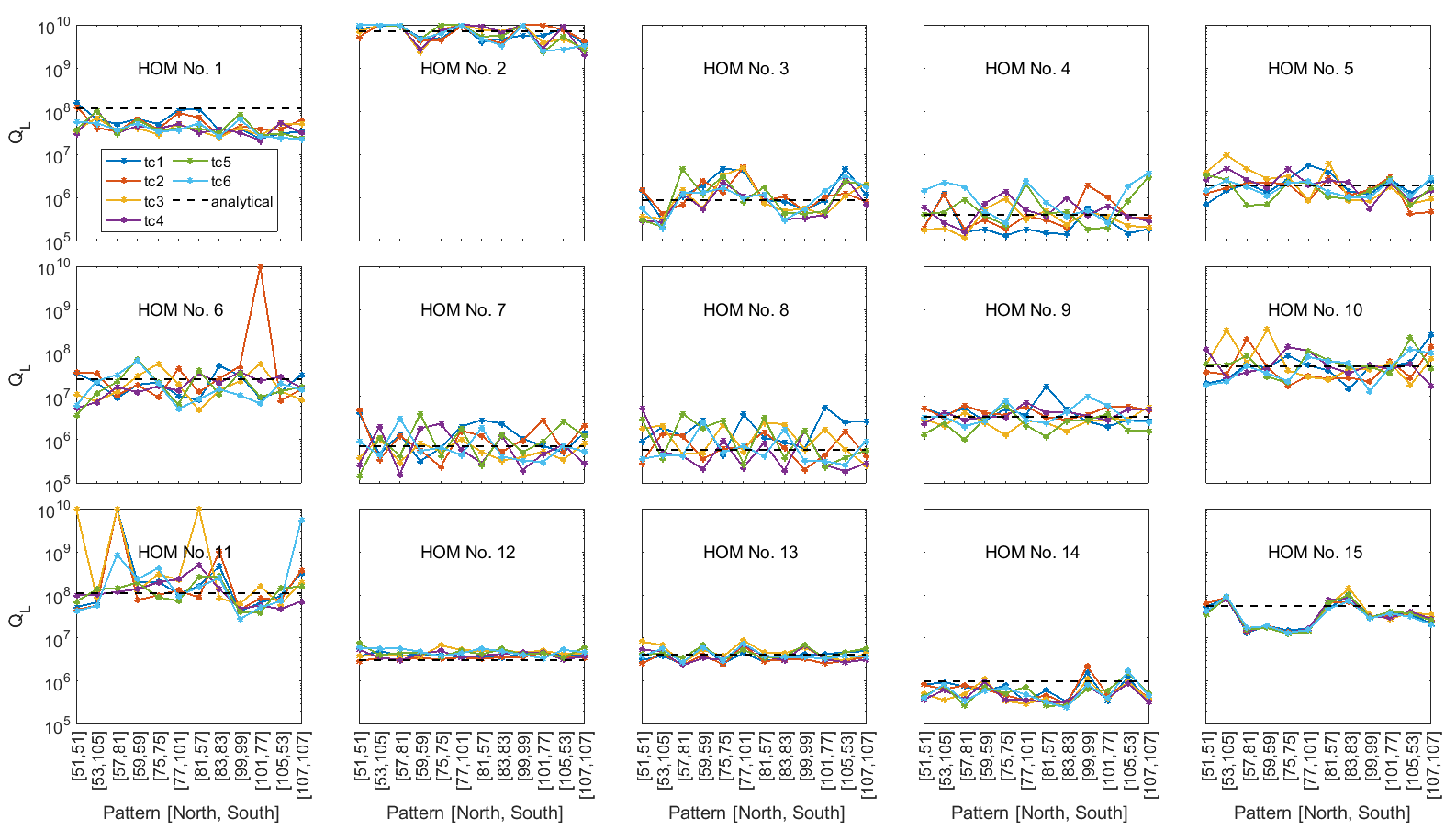}}
	\caption{Simulation results for the 12 pattern and 6 timing combinations (tc). }
	\label{fig:SimRes1} 
\end{figure*}

\subsection{Pattern and timing dependence}
\begin{figure}
    \begin{tabular}{c}
		\def\stackalignment{l}
		\topinset{\bfseries(a)}{\includegraphics[width=3.0in]{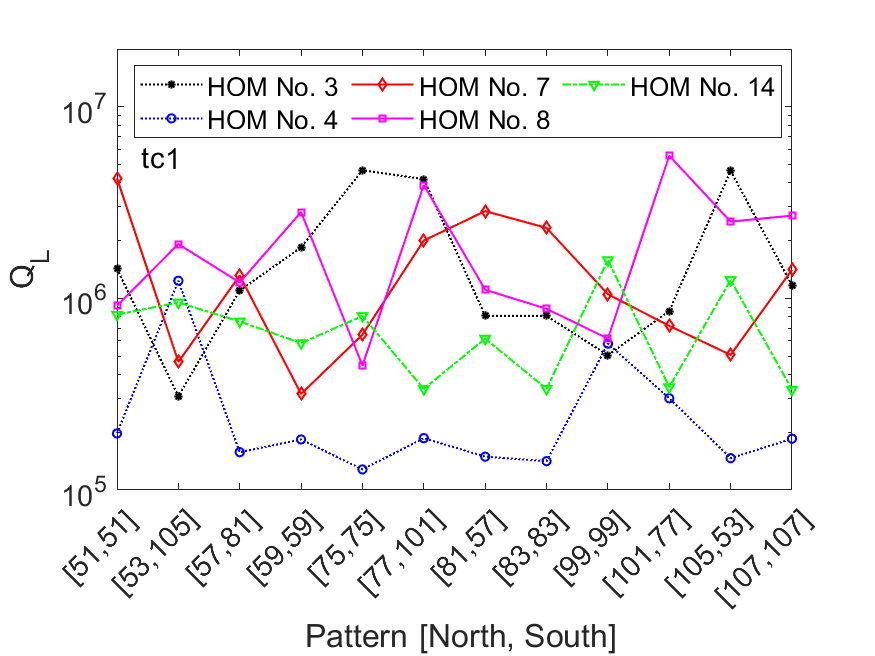}}{0.55in}{2.0in} \\
		\def\stackalignment{l}
		\topinset{\bfseries(b)}{\includegraphics[width=3.0in]{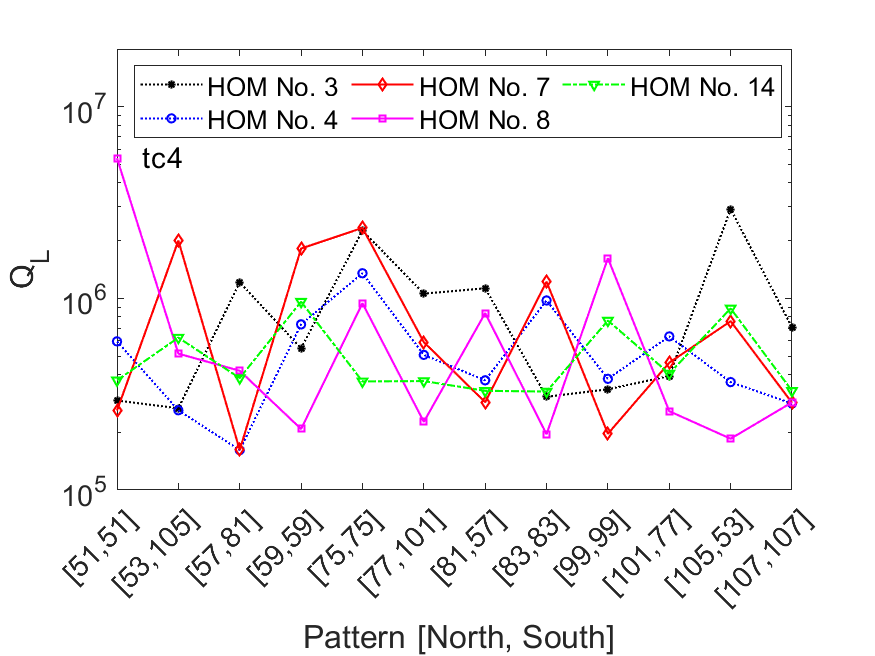}}{0.55in}{2.0in} \\    
%		\def\stackalignment{l}
%		\topinset{\bfseries(b)}{\includegraphics[width=3.0in]{fig/002_HOM_3_4_7_8_14tc2}}{0.55in}{2.0in} \\	
%		\def\stackalignment{l}
%		\topinset{\bfseries(c)}{\includegraphics[width=3.0in]{fig/002_HOM_3_4_7_8_14tc3}}{0.55in}{2.0in} 
%		\def\stackalignment{l}
%		\topinset{\bfseries(d)}{\includegraphics[width=3.0in]{fig/002_HOM_3_4_7_8_14tc4}}{0.55in}{2.0in} \\  
%  	\def\stackalignment{l}
%		\topinset{\bfseries(e)}{\includegraphics[width=3.0in]{fig/002_HOM_3_4_7_8_14tc5}}{0.55in}{2.0in} 
%		\def\stackalignment{l}
%		\topinset{\bfseries(f)}{\includegraphics[width=3.0in]{fig/002_HOM_3_4_7_8_14tc6}}{0.55in}{2.0in} \\
	\end{tabular}
	\caption{Required $Q_L$ scan results for 5 worst HOMs for timing combination 1 (a) and 4 (b).}
	\label{fig:5HOMs} 
\end{figure}

Simulation results show HOM No. 3, 4, 7, 8, and 14 are the most critical modes. It can be seen the they are pattern and timing dependent as shown in Fig.~\ref{fig:5HOMs}. In sub-figure (a), which is timing combination 1, the HOM No.~4 is the critical model in most cases, while in timing combination 4 in sub-figure (b) other modes becomes critical. This indicates that both filling patterns and timing combinations have a significant impact on BBU instability. 

%\begin{figure*}
%\scalebox{0.42} [0.42]%{\includegraphics{fig/001_TimingScan5_3WorstModes}}
%	\caption{Overlapped results for 3 worst HOMs. }
%	\label{fig:SimRes1} 
%\end{figure*}

\subsection{HOM voltage oscillation and BBU instability}
The voltage in one cavity impacts voltage in other cavities through bunch offsets $x$ and henceforth beam loading $dV_{\mathrm{HOM}}$. For example, the kick $\Delta x_{cav 1}'$ received by the bunch in the first cavity can be given as

\begin{equation}
   \Delta x_{cav 1}' = \frac{V_{\mathrm{HOM},~cav 1,~im}}{V_{beam}}
   \label{eq:x'1}
\end{equation}

\noindent where ${V_{\mathrm{HOM},~cav 1,~im}}$ is the imaginary part of the HOM voltage as the kick is from the magnetic field and $V_{beam}$ is the beam voltage, $pc/e$. This kick adds an offset in the second cavity 

\begin{equation}
   x_{cav 2} = M_{11}x_{cav 1} + M_{12}\Delta x_{cav 1}'
   \label{eq:x2}
\end{equation}

\noindent with $M_{11}$ and $M_{12}$ being the elements transfer matrix between the first and second cavity. This would in turn impact the beam loading in the second cavity $dV_{\mathrm{HOM},~cav 2}$ which can be given by 

\begin{equation}
    dV_{\mathrm{HOM},~cav 2} = \frac{(2\pi f_{\mathrm{HOM}})^2_{H}}{2c} q_{b} \left( \frac{R}{Q} \right)_{H} x_{cav 2}\\
    \label{eq:dv1}
\end{equation}

\noindent where $\omega_{\mathrm{HOM}} = 2\pi f_{\mathrm{HOM}}$ with $f_{HOM}$ being the HOM frequency, $q_{b}$ is the bunch charge and $\left(\frac{R}{Q}\right)_{H}$ is geometric shunt impedance of the HOM. Inserting Eq.~\ref{eq:x'1} and Eq.~\ref{eq:x2} in to Eq.~\ref{eq:dv1} gives

\begin{equation}
\begin{array}{l}
    dV_{\mathrm{HOM},~cav 2} = \\
    \frac{\omega_{\mathrm{HOM}}^2}{2c} q_{b} \left( \frac{R}{Q} \right)_{H} \left[M_{11}x_{cav 1} + M_{12} \left( \frac{V_{\mathrm{HOM},~cav 1}}{V_{beam}} \right) \right].
    \end{array}
    \label{eq:dv2}
\end{equation}

\noindent Eq.~\ref{eq:dv2} clearly shows how the HOM voltage in the first cavity can impact the second cavity. 

\begin{figure}
	\begin{tabular}{c}
		\def\stackalignment{l}
		\topinset{\bfseries(a)}{\includegraphics[width=3.5in]{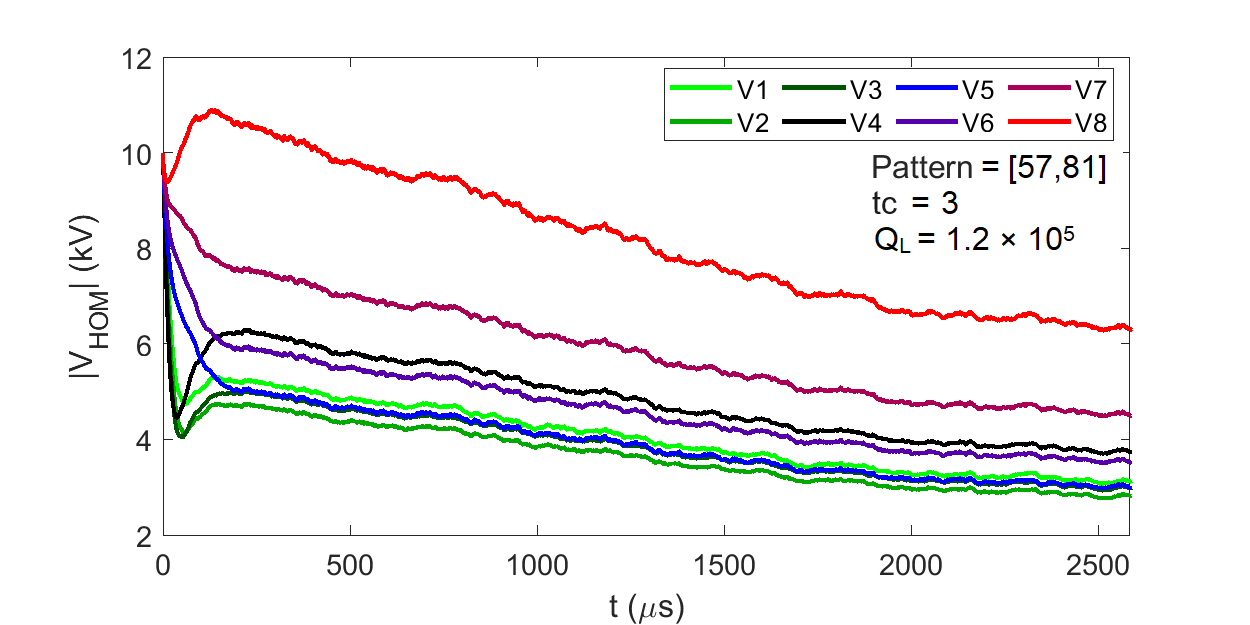}}{0.2in}{1.3in} \\
		\def\stackalignment{l}
		\topinset{\bfseries(b)}{\includegraphics[width=3.5in]{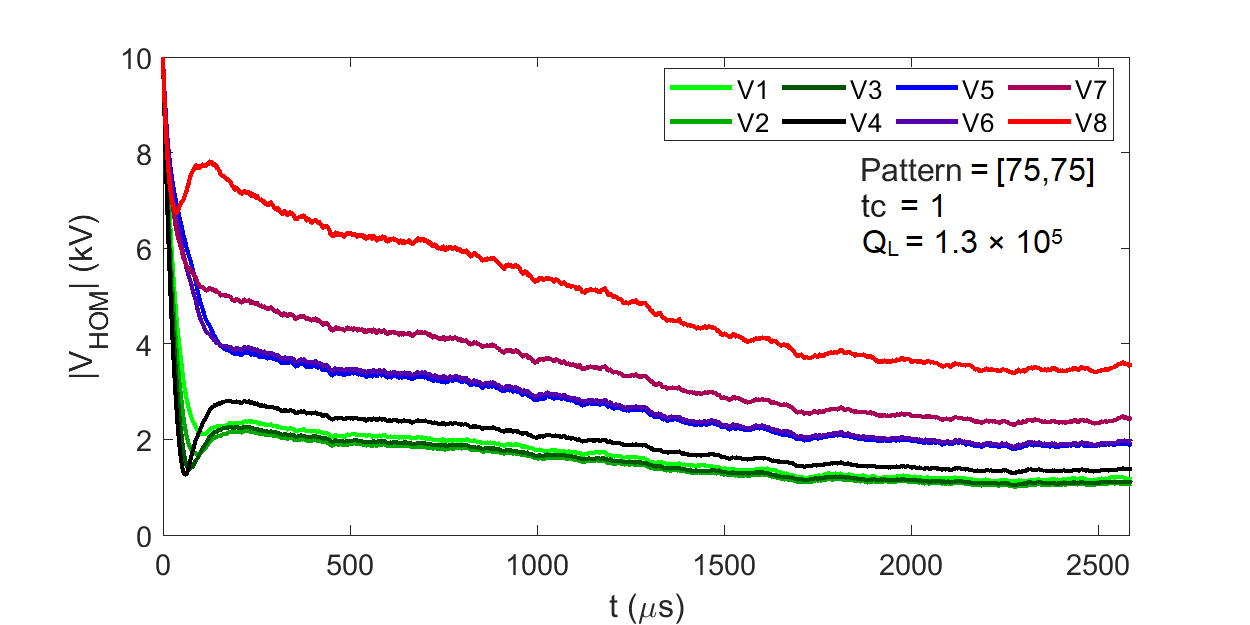}}{0.2in}{1.3in} \\	
		\def\stackalignment{l}
		\topinset{\bfseries(c)}{\includegraphics[width=3.5in]{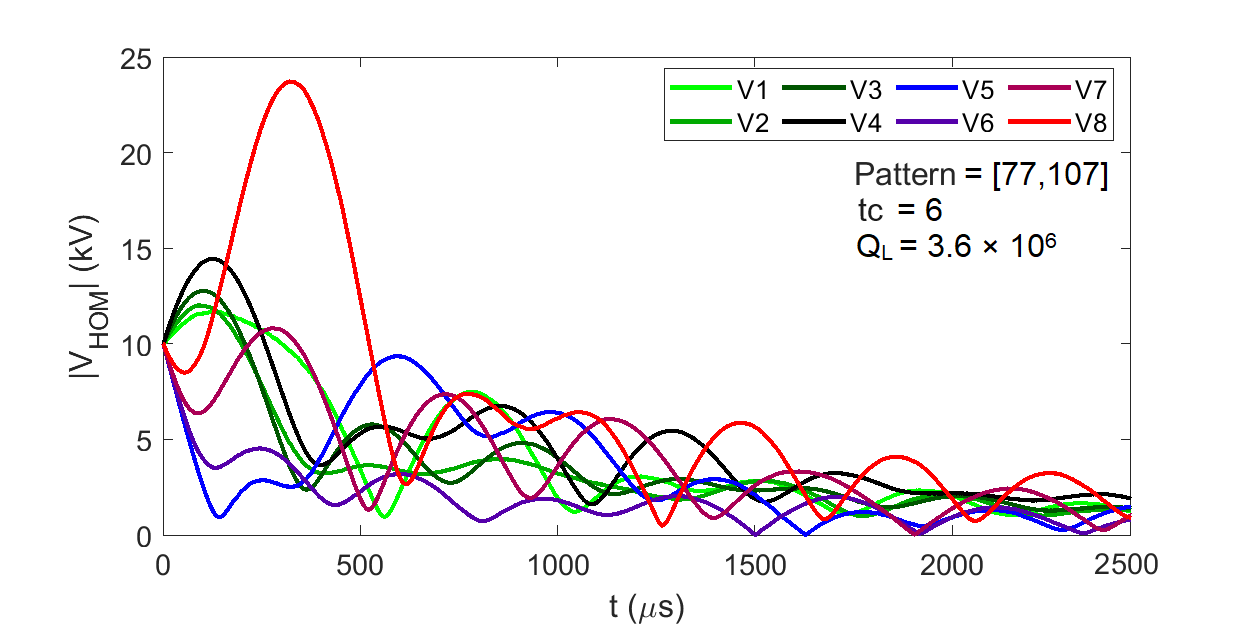}}{0.2in}{1.3in} \\
		\def\stackalignment{l}
		\topinset{\bfseries(d)}{\includegraphics[width=3.5in]{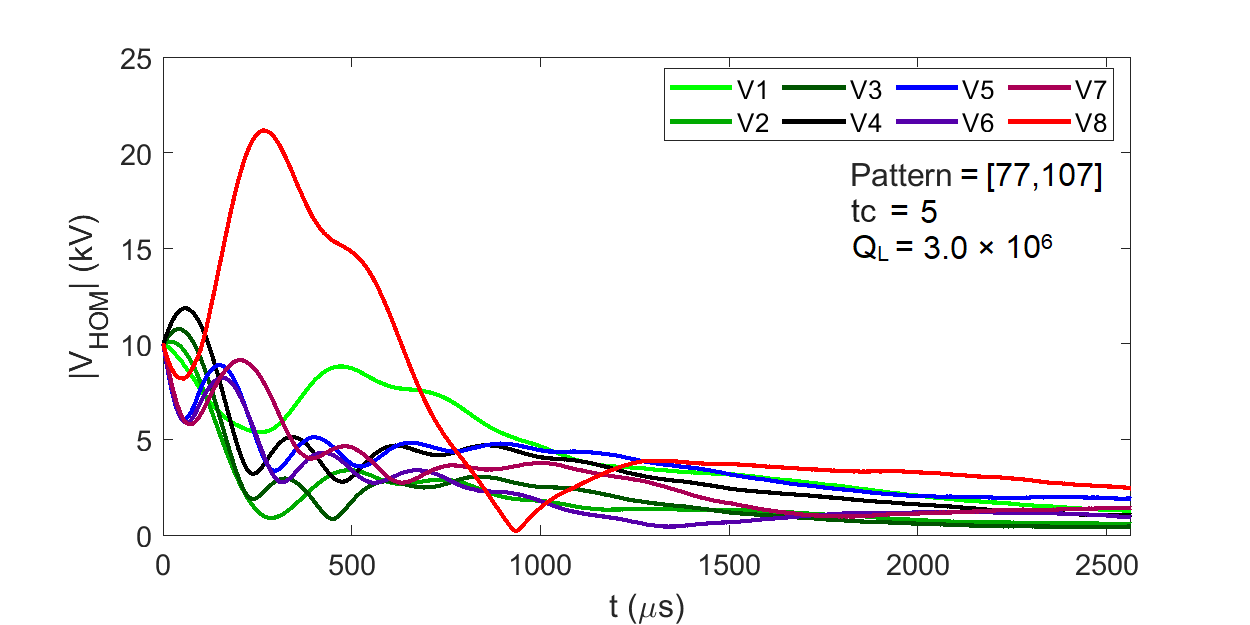}}{0.2in}{1.3in} \\  
	\end{tabular}
	\caption{Cavity voltages of highest and lowest $Q_L$s of HOM No.4.}
	\label{fig:VHOM}
\end{figure}

All 8 cavities are interconnected through bunch offset. The behavior of HOM voltages can be likened to a set of interconnected balls, with the bunch offset acting as a spring that transfers oscillation from one cavity to another. As a result, there is a synchronization throughout the system, which causes HOM voltages to exhibit similar fluctuations and trends, as demonstrated in Figure~\ref{fig:VHOM}, particularly in subfigures (a) and (b), where the cavities share noticeable small voltage oscillations.

In Fig.~\ref{fig:VHOM}, the HOM voltage behaviors in low-$Q_L$ cases shown in sub-figures (a) and (b) are different from those of high-$Q_L$ cases shown in (c) and (d). In low-$Q_L$, the voltages are synchronized much faster indicating stronger coupling between cavities. Even minor fluctuations are commonly shared by all cavities. The high-$Q_L$ cases have more oscillations, longer synchronization and settling time, and significant lagging in oscillations, which indicates weaker coupling between cavities. In summary, the $Q_L$ is inversely proportional to the coupling between cavities.

\begin{figure}
\scalebox{0.4} [0.4]{\includegraphics{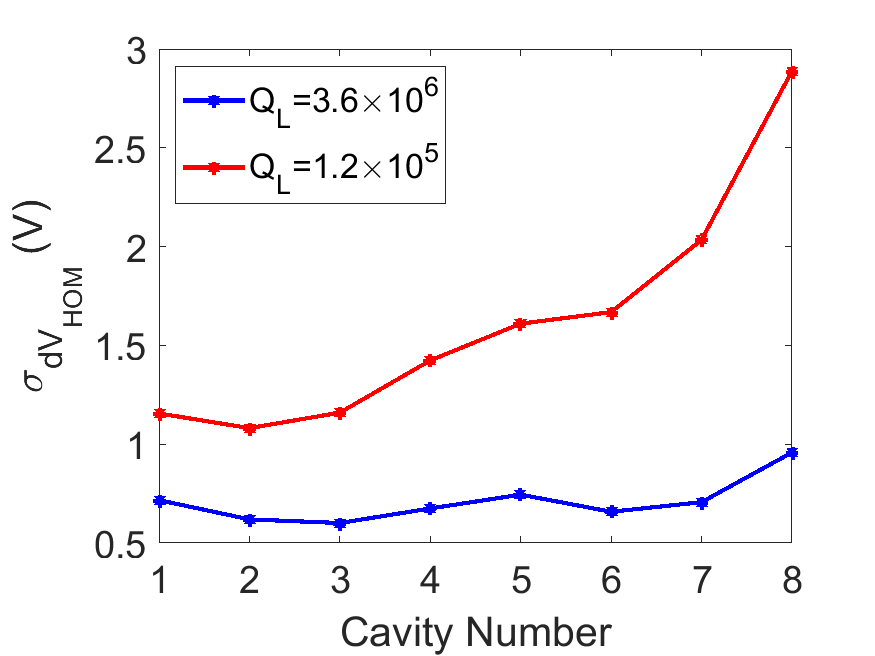}}
	\caption{$\sigma_{dV_{\mathrm{HOM}}}$ of maximum and minimum $Q_L$s of HOM No. 4 of 8 cavities. }
	\label{fig:SimRes2} 
\end{figure}

The coupling between cavities is proportional to beam loading, and stronger beam loading leads to stronger cavity coupling.  The standard deviation of beam loading ($\sigma_{dV_{\mathrm{HOM}}}$) over 1 turn indicates the strength of beam loading. Beam loading is stronger in $Q_{L} = 1.2 \times 10^5$ compared to $Q_{L} = 3.6 \times 10^6$, as seen in Figs.~\ref{fig:SimRes1} and \ref{fig:SimRes2}. With stronger beam loading, oscillations propagate more easily to other cavities, necessitating a lower $Q_L$ for faster HOM damping.

\begin{figure*}
\scalebox{0.35} [0.35]{\includegraphics{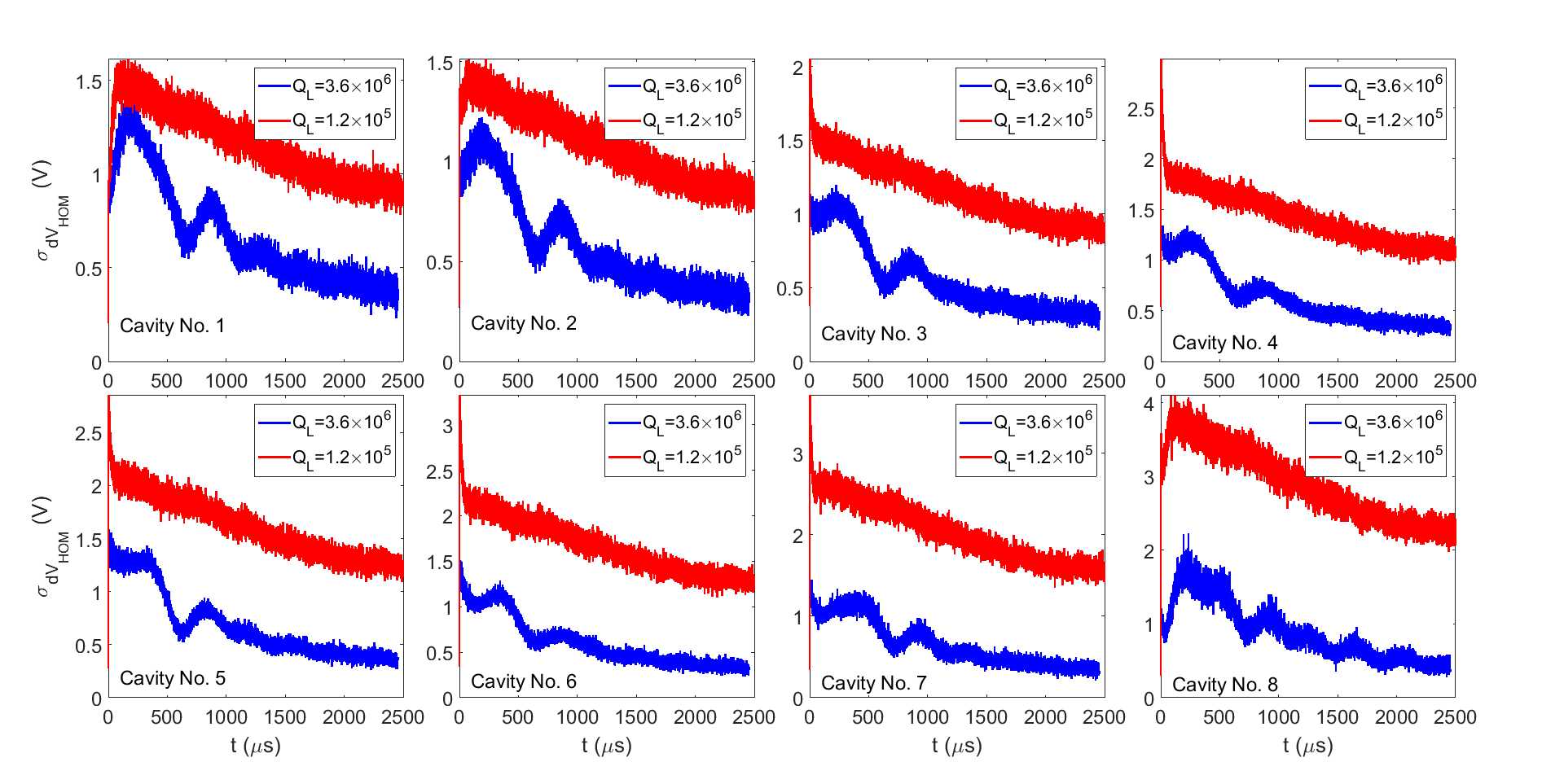}}
	\caption{$\sigma_{dV_{\mathrm{HOM}}}$ of max and min $Q_L$s of HOM No. 4 as a function of time. }
	\label{fig:SimRes1} 
\end{figure*}

When the beam loading is strong, oscillations are more easily propagated to other cavities, so the $Q_{L}$ needs to be lower so HOM can be dampened faster. It can be seen in Fig.~\ref{fig:SimRes2} that later cavities tend to have stronger beam loading and cavity No. 8 has the highest. This difference in beam loading causes the cavities to have different final settling voltages. For example, cavity No.~8 always highest beam loading as shown in Fig.~\ref{fig:SimRes2}, which caused it to have the highest settling voltage as shown in Fig.~\ref{fig:VHOM}. This indicates there is more HOM build-up in later cavities that would require more damping and care should be given to them.

The optimal patterns and timings can lower the beam loading significantly, which can lower coupling between cavities and hence reduce the propagation of HOM through cavities. This can suppress BBU instability and allow required $Q_L$ to be higher (i.e. less HOM damping is required). 

%\section{Differential beam loading}
%The final settling voltages of the cavity are dependent on the position of the cavity. Here the settling voltage refers to the cavity voltage when there are no more oscillations and voltages are mostly flat. The earlier cavities have lower settling voltages than the later ones.

\section{Threshold Current Estimation}
PERLE HOM (Higher Order Modes) couplers have been designed by Barbagallo \textit{et al.}~\cite{Carmelo2022ERL} to mitigate unwanted modes while not affecting the fundamental one. The loaded Q-factors ($Q_L$) of the HOMs were effectively reduced below critical levels. Still, it remains essential to ascertain the threshold current when all modes are operative simultaneously. We further enhanced our numerical model to estimate threshold current when all modes are activated. Simulations were carried out with lowered $Q_L$s. It's notable that these couplers disrupted the transverse symmetry, which means the $Q_L$ and $R/Q$ for the vertical and horizontal modes are different. In the simulations, a total of 30 modes are considered, half of which are horizontal and the other half are vertical. 

\subsection{Simulations results}
\begin{figure}
\scalebox{0.35} [0.35]{\includegraphics{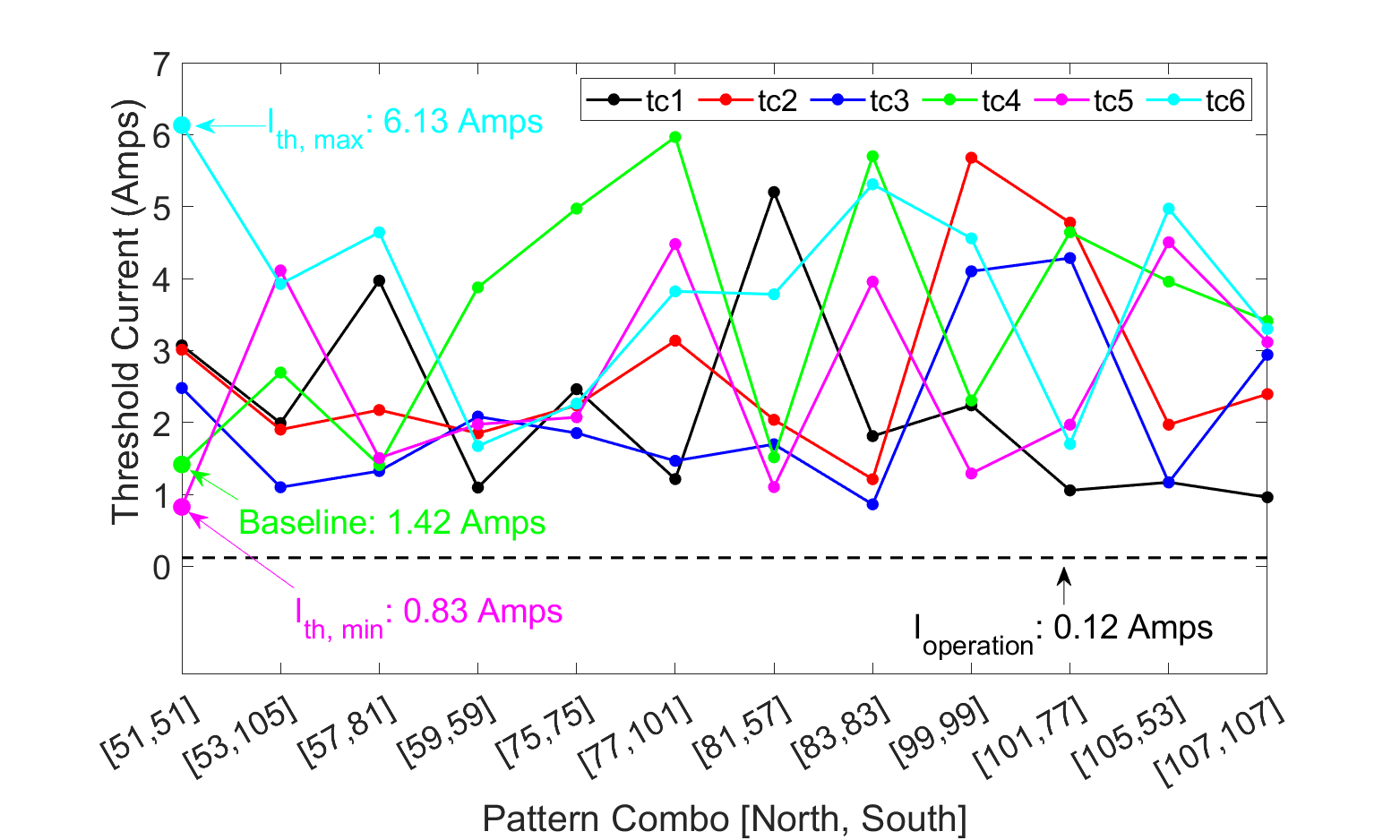}}
	\caption{Threshold current scan results with all modes activated. Each curve represent different timing combination (tc). }
	\label{fig:SimRes3} 
\end{figure}

Simulations were conducted encompassing 12 pattern and 6 timing combinations, the results of which are depicted in Fig.~\ref{fig:SimRes3}. The existing design of PERLE employs pattern combination [51, 51] and timing combination No.4, yielding a threshold current of 1.42 Amps. Impressively, this value is nearly 12-fold greater than PERLE's operational current of 0.12 Amps.

The threshold current is found to be highly sensitive to both the bunch pattern and the timing. By adjusting the bunch timing alone, we have been able to achieve a maximum threshold current of 6.13 Amps or, conversely, a minimum of 0.83 Amps. Regardless of these variations, it's noteworthy that the threshold current remains significantly higher than the operational current of 0.12 Amps in all scenarios. This outcome indicates that the HOM couplers are effectively damping HOMs, thus ensuring stable operation even at high threshold currents.

\subsection{Dominant modes}
\begin{figure}
	\begin{tabular}{c}
		\def\stackalignment{l}
		\topinset{\bfseries(a)}{\includegraphics[width=3.5in]{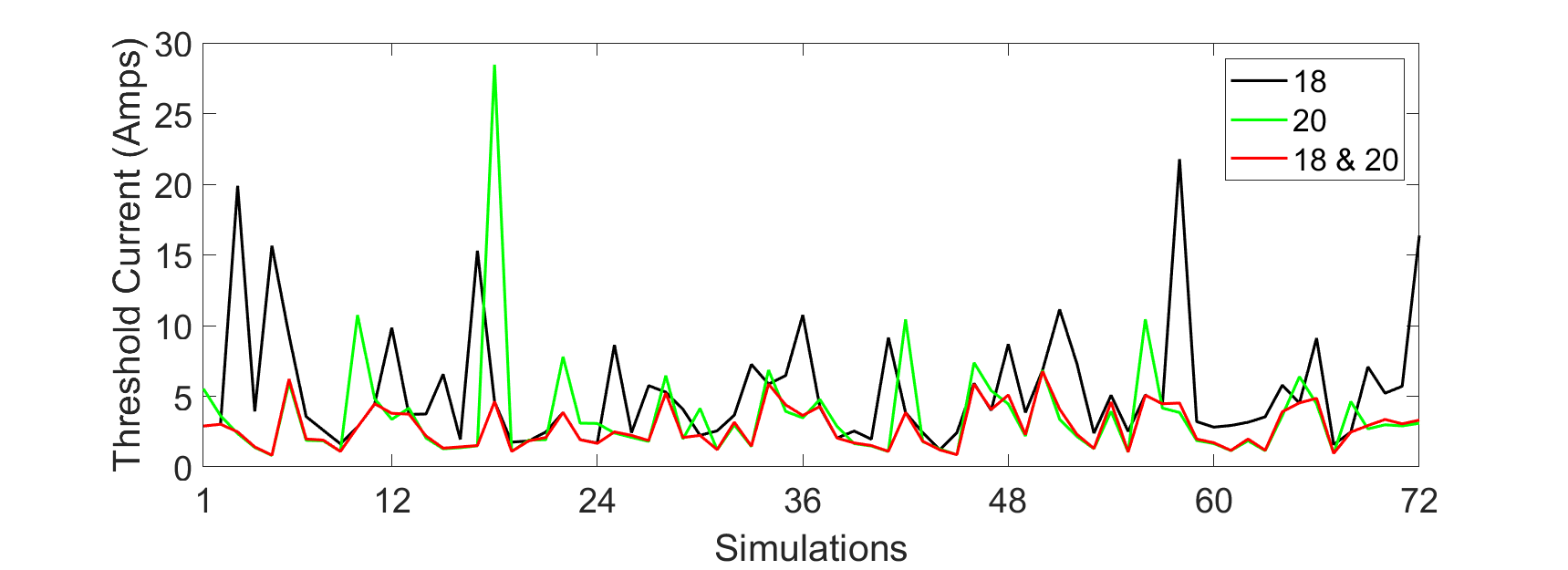}}{0.2in}{1.3in} \\
		\def\stackalignment{l}
		\topinset{\bfseries(b)}{\includegraphics[width=3.5in]{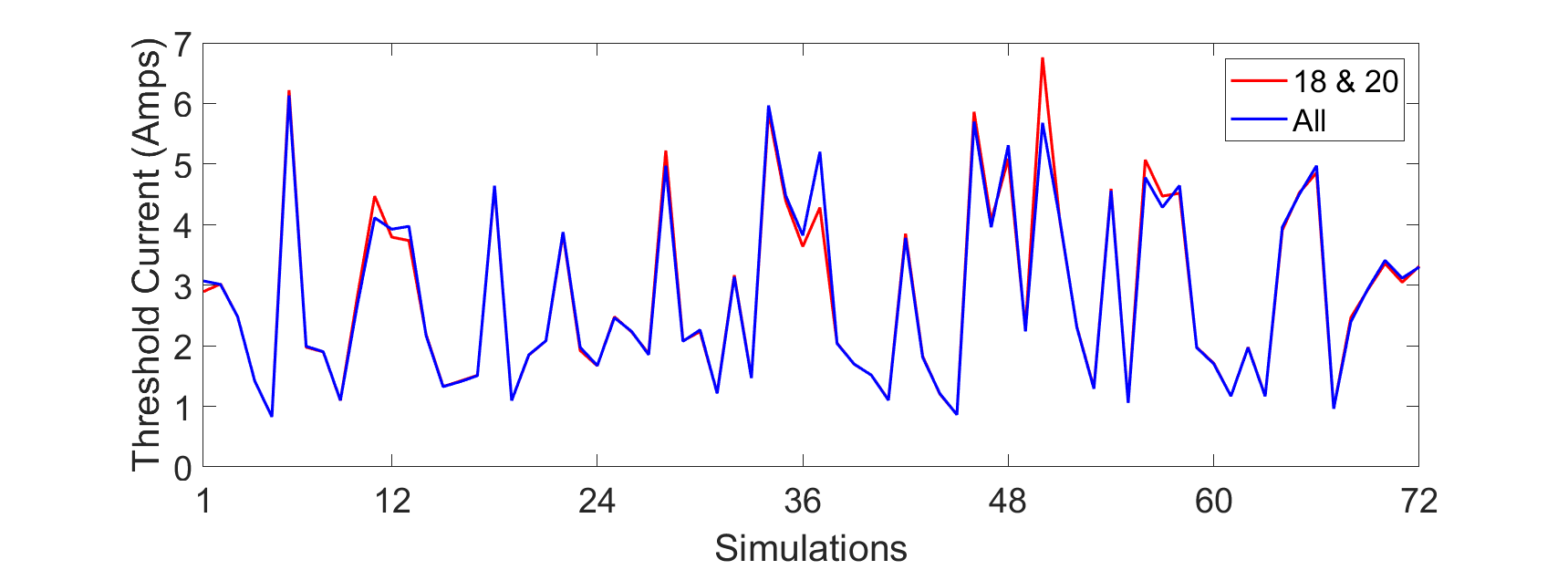}}{0.2in}{1.3in} \\	 
	\end{tabular}
	\caption{Threshold current results: (a) comparison of activating all modes (blue) versus activating modes 18 and 20 (red) only; (b) comparison of activating modes 18 and 20 (red) versus activating a single mode 18 (black) and mode 20 (green).}
	\label{fig:VHOM2}
\end{figure}

In our simulations, we observed the threshold currents were mostly set by mode No.~18 or 20, which are both vertical modes. To investigate which mode is the most dominant mode, we also simulated the situation where only a selected few modes are activated and the results are given in Fig.~\ref{fig:VHOM2} for 12 filling patterns and 6 bunch timings. In the subfigure (a), the red curve indicates the case where only modes 18 and 20 are activated and the blue curve is when all 30 modes are activated. It shows the threshold current is mostly dictated by the modes 18 and 20. The subfigure (b) shows mode 20 is slightly more dominant than 18. 

\subsection{Frequency Jitters}

\begin{figure*}
\scalebox{0.45} [0.45]{\includegraphics{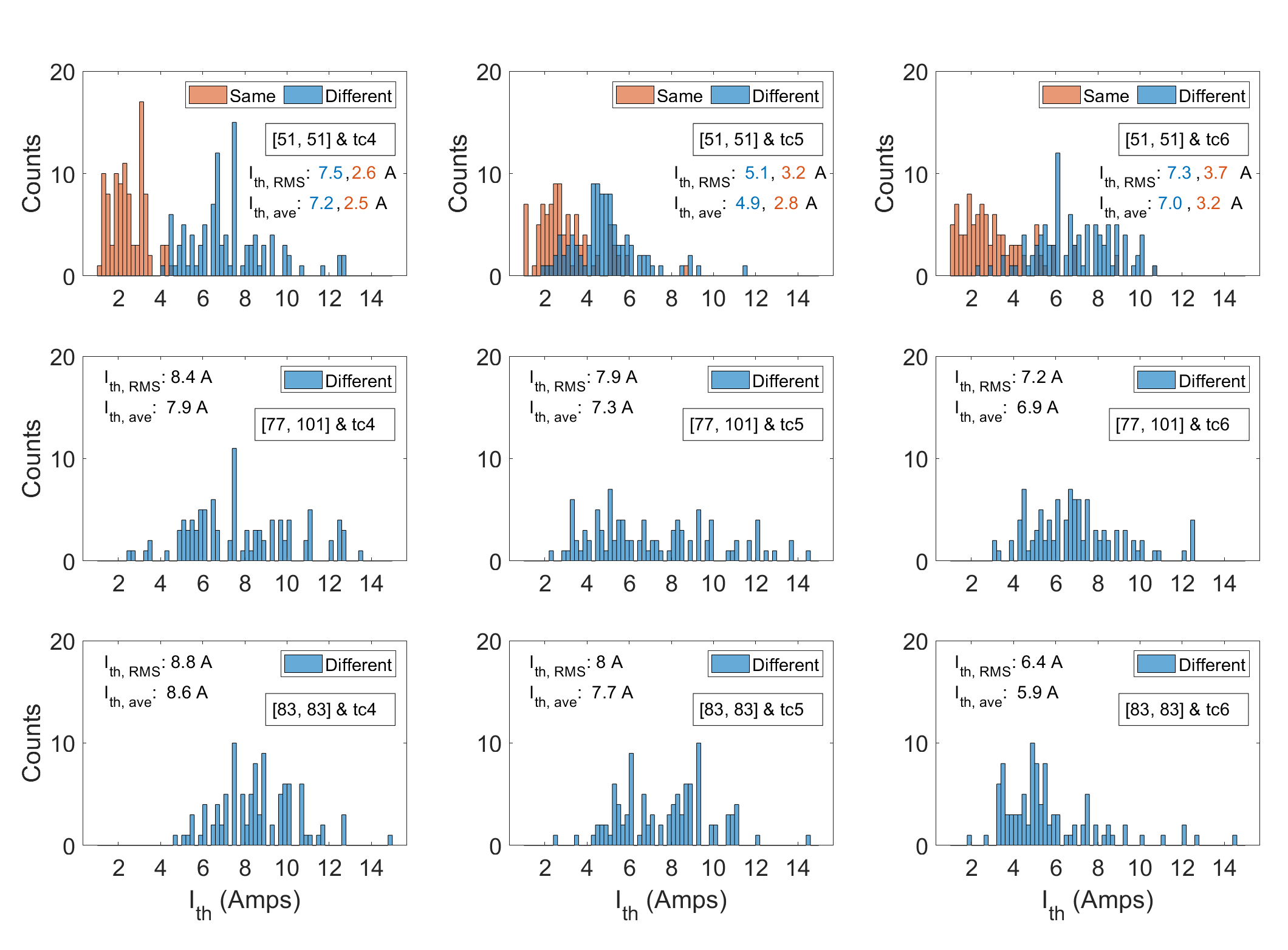}}
	\caption{Simulation with varied filling patterns and bunch timings with jitters of $\sigma_{f_{\mathrm{HOM}}}/{f_{\mathrm{HOM}}} = 0.001$. The orange bins indicates the case when cavities have the same random HOM frequency, while blue indicates they have different HOM frequencies.}
	\label{fig:JitSim1} 
\end{figure*}

The HOM spectrum of the manufactured cavities can vary from the design. 
As can be seen from Eq.~\ref{eq:IthAna}, the threshold current is sensitive to the HOM frequency. Slight changes in the frequency can vary the threshold current significantly, as show in the Ref.~\cite{Setiniyaz2021}. Therefore, relative RMS jitters of $\sigma_{f_{\mathrm{HOM}}}/{f_{\mathrm{HOM}}} = 0.001$ were introduced to the simulations assuming Gaussian distribution for 3 different filling patterns and 3 timings and results are given in the Fig.~\ref{fig:JitSim1}. The orange distributions indicate the case where all the cavities have the same randomly assigned frequency. The blue ones are when the 8 cavities have different randomly assigned frequency. It can be seen that (1) similar ranges of threshold currents (few to more than 10 Amps) were observed between different filling patterns and timings; (2) the threshold current is significantly higher when cavities have different frequencies; and (3) the lowest threshold currents predicted in simulations are few Amps, which is an order higher than the PERLE design requirement of 0.12 Amps. 

\begin{figure}
	\begin{tabular}{c}
		\def\stackalignment{l}
		\topinset{\bfseries(a)}{\includegraphics[width=3.2in]{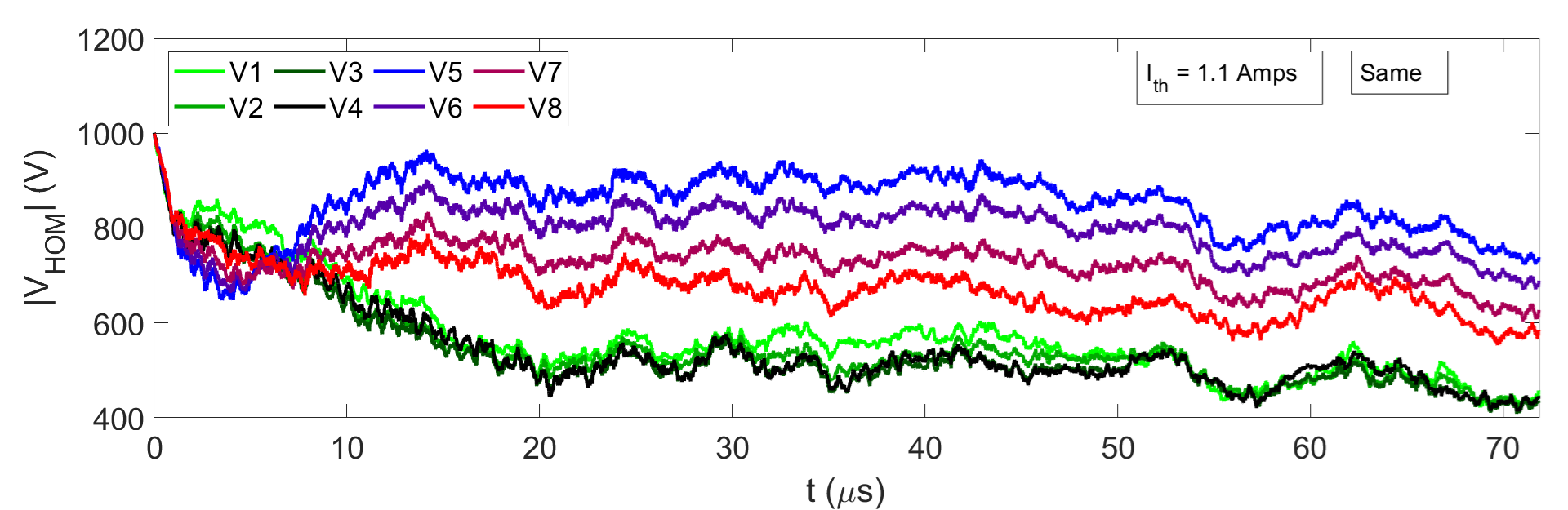}}{0.27in}{2.9in} \\
		\def\stackalignment{l}
		\topinset{\bfseries(b)}{\includegraphics[width=3.2in]{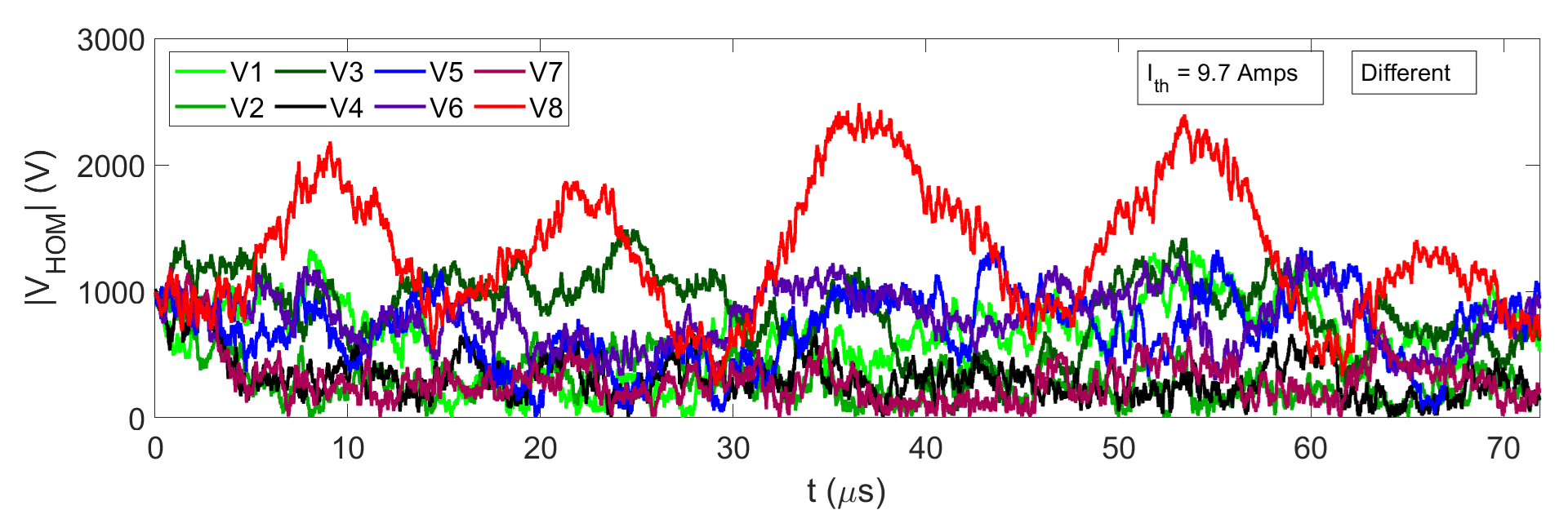}}{0.27in}{2.9in} \\	 
	\end{tabular}
	\caption{HOM voltages when the 8 cavities have same (a) and different (b) HOM frequencies. In (a) cavities are coupled and synchronized as they share same HOM frequency. In (b), the coupling between cavities are broken as they have different HOM frequencies.}
	\label{fig:VHOM3}
\end{figure}

It can be seen from Fig.~\ref{fig:VHOM3} (a), when the cavities have same HOM frequencies, they can form resonances and oscillate together. This can amplify BBU and result in a low threshold current. When HOM frequencies are different as in sub-figure (b), the resonance is broken and the threshold current increased by nearly an order of magnitude in this case. 

We also varied the relative RMS jitters $\sigma_{f_{\mathrm{HOM}}}/{f_{\mathrm{HOM}}}$ and no significant difference is observed. This is because the threshold current is quasi-periodic over HOM frequency~\cite{Setiniyaz2021}, which can also be seen from the Eq.~\ref{eq:IthAna}. As PERLE revolution time is around 0.2~$\mu$s, the half period of threshold is approximately 2.5~MHz. As the HOM frequencies are on the order of GHz, relative RMS jitters of 0.001 would sufficiently cover the 1 threshold period.

\begin{figure}
\scalebox{0.45} [0.45]{\includegraphics{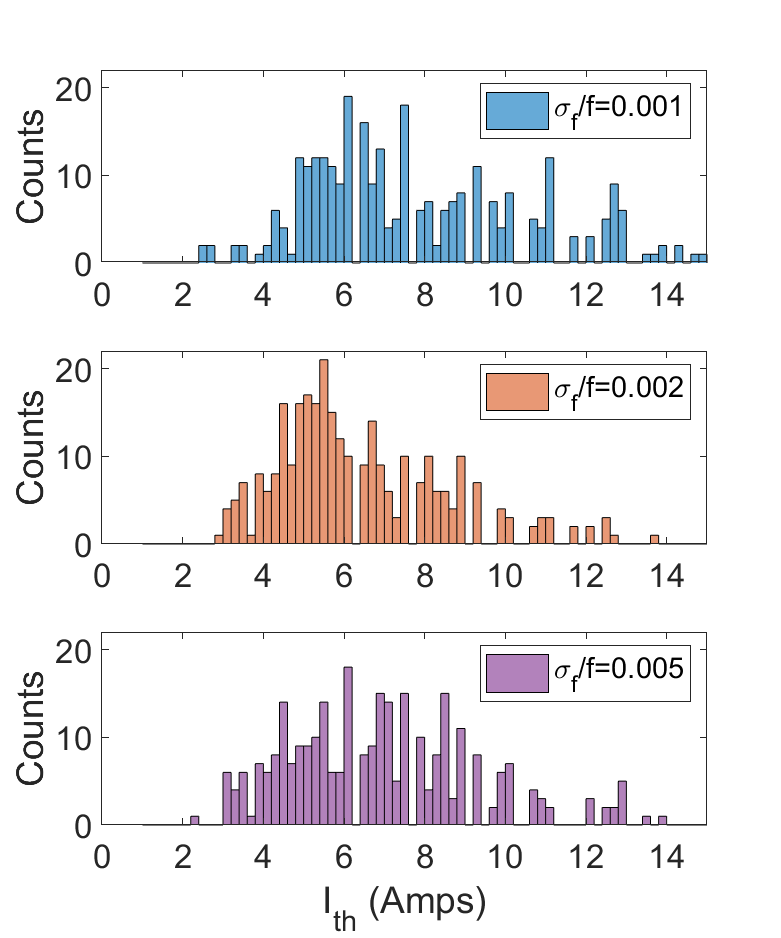}}
	\caption{Threshold currents with different relative RMS jitters. }
	\label{fig:JitSim2} 
\end{figure}

\section{Threshold current estimated in analytical model and Bmad}

\begin{figure}
\scalebox{0.35} [0.35]{\includegraphics{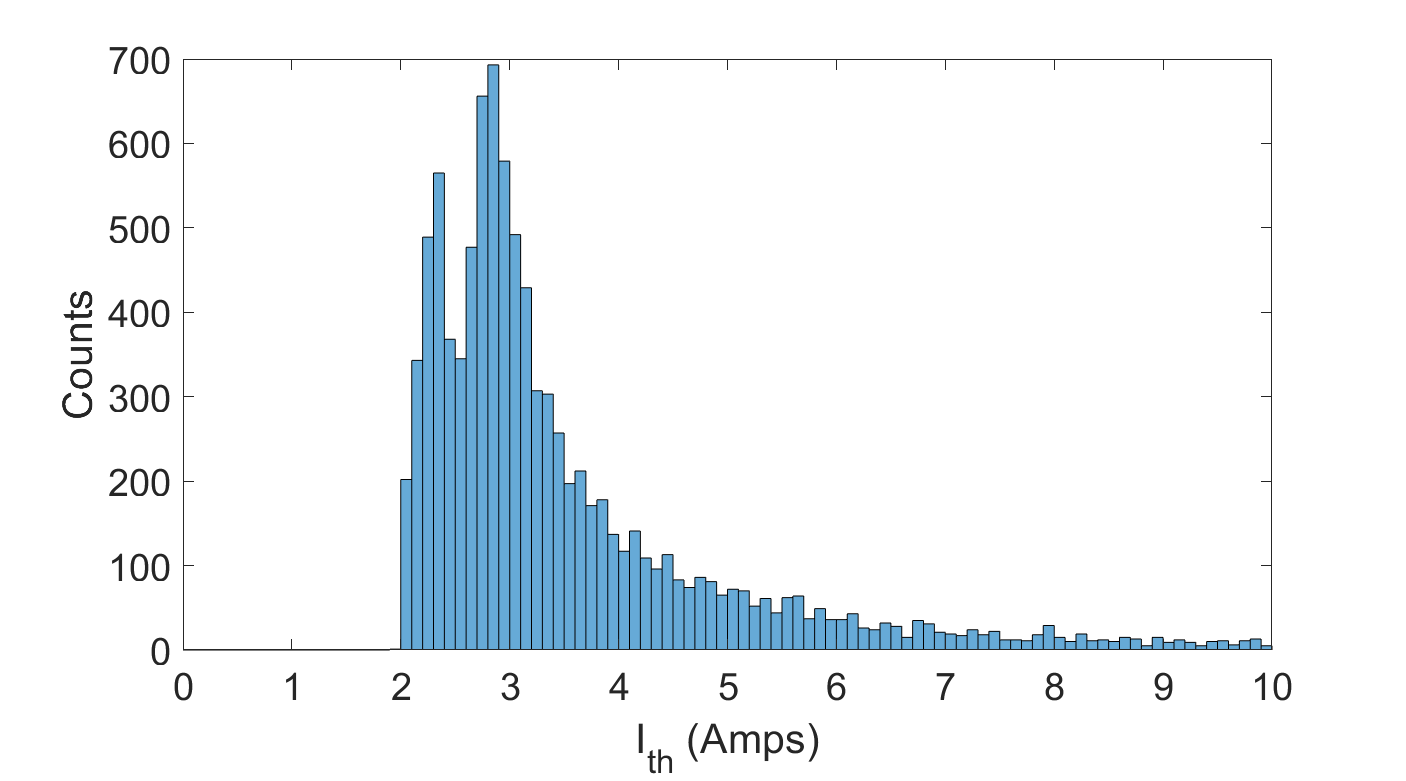}}
	\caption{Threshold current obtained in analytical model in Eq.~\ref{eq:IthAna} with jitters of $\sigma_{f_{\mathrm{HOM}}}/{f_{\mathrm{HOM}}} = 0.001$. }
	\label{fig:JitAnaRes} 
\end{figure}

So far, we have only reported threshold current results from our in-house BBU tracking code. To crosscheck these results, we estimated threshold currents using both the analytical model and Bmad~\cite{Bmad}. Similar to our previous approach, we introduced relative RMS jitters $\sigma_{f_{\mathrm{HOM}}}/{f_{\mathrm{HOM}}} = 0.001$ into the analytical model, described in Eq.~\ref{eq:IthAna}. The resulted threshold currents are shown in the Fig.~\ref{fig:JitAnaRes}. We observe that the minimum threshold current is around 2~Amps, which is consistent with earlier simulations. When the RMS jitters were varied to $\sigma_{f_{\mathrm{HOM}}}/{f_{\mathrm{HOM}}}$ = 0.002 and 0.005, no significant difference are observed in the threshold current distributions, which is also similar to the results of the earlier simulations. The distribution of the threshold currents is different to of the simulations, this is due to the fact that the analytical model is different from the simulations in the analytical model doesn't account for the phases of bunches and interaction between cavities etc. 

To perform BBU studies in Bmad, the original PERLE 2.0 lattice was converted from OptiMX to Bmad. Once the lattices were converted, it was necessary to rematch the beamlines, as the approximations made for cavity edge focusing differ by a small amount. Once rematched, the sections were concatenated together, and Bmad's multipass functionality was applied, whereby beamline elements which are common to multiple passes of different energies are identified as such, and the appropriate calculations are performed to insure consistency between the different energies in each common element. The optics and energy recovery was checked in Bmad, and compared against the original design code.

\begin{figure}
\scalebox{0.35} [0.35]{\includegraphics{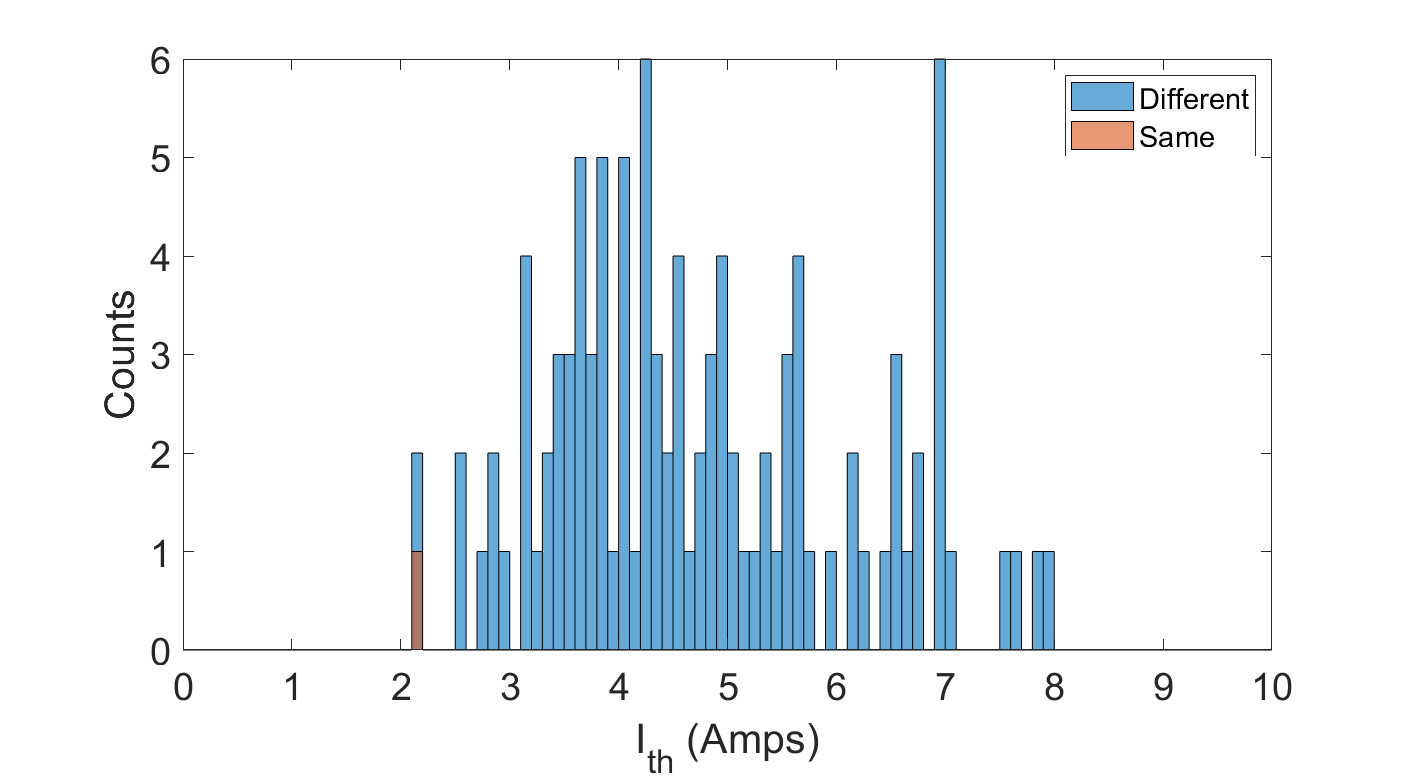}}
	\caption{Threshold currents obtained in Bmad with jitters of $\sigma_{f_{\mathrm{HOM}}}/{f_{\mathrm{HOM}}} = 0.001$. }
	\label{fig:JitBmadRes} 
\end{figure}

The threshold current results are given in the Fig.~\ref{fig:JitBmadRes} for the baseline PERLE filling pattern combination [51, 51] and timing combination No.4. When the cavities had same HOM parameters, the threshold current was at around lowest value of 2.1~A. When random jitters of $\sigma_{f_{\mathrm{HOM}}}/{f_{\mathrm{HOM}}} = 0.001$ introduced, the threshold current increased. We can see the results are consistent with the predictions of in-house tracking code and analytical model.

\section{Conclusion}
In this work, we explored all possible filling patterns and bunch timing combinations for PERLE with its current constraints.
We built an 8-cavity PERLE BBU tracking model and numerically estimated the damping requirements for HOMs, finding strong agreement with the analytical model. As the numerical model is more sophisticated, it was able to incorporate the impacts of filling patterns, bunch timings, and HOM frequency difference in cavities, providing deeper insights into the behavior of the 8-cavity ERL system. 

In our simulations, we observed that when cavities share same HOM frequencies, the become interconnected through bunch offset and beam loading, which led to the synchronization and propagation of HOM voltages across the cavities. 
However, slight variation in HOM frequencies (by $\sigma_{f_{\mathrm{HOM}}}/{f_{\mathrm{HOM}}} = 0.001$) can disrupt this synchronization, mitigate BBU instability, and increase the threshold current to several Amps.

Our analysis indicates that bunch timings are as influential as filling patterns. By optimizing these elements, we can diminish beam loading and interaction between cavities, reducing the spread of HOM voltages between cavities. This, in turn, helps control BBU instability and raises the threshold current. 

We used an analytical model and two BBU tracking codes to estimate threshold current of PERLE at least to be around 2~Amps. This is 17 times larger than the required operation current of 0.12~Amp. The results show frequency jitters of $\sigma_{f_{\mathrm{HOM}}}/{f_{\mathrm{HOM}}} = 0.001$ are sufficient to increase the threshold current by an order of magnitude. 

Among these factors, the bunch timing and filling pattern can be adjusted by carefully designing the beamline lattice. In contrast, the HOM frequency variations are fixed once cavities are manufactured. Therefore, integrating a mechanism to adjust bunch timing and filling patterns to the multi-turn ERLs is crucial for managing BBU instability. 

\section{ACKNOWLEDGEMENTS}
The authors extend their sincere thanks to Dr. Graeme Burt for his invaluable suggestions and insights. 
A special thanks to Julien Michaud and the support provided by the PERLE collaboration. We also express our gratitude to the Bmad development team for their assistance. The studies presented have been funded by STFC Grants No. ST/P002056/1 and ST/V001612/1 under the Cockcroft Institute Core Grants.
Work at Jefferson Lab has been supported by the U.S. Department of Energy, Office of Science, Office of Nuclear Physics under contracts DE-AC05-06OR23177. % and DE-SC0012704.

%\section{APPENDIX}
%\bibliographystyle{unsrt}

\end{document}